\documentclass[english,12pt,aps,prd,a4paper,preprintnumbers,floatfix,nofootinbib,showpacs,superscriptaddress, notitlepage]{revtex4-1}
 \pdfoutput=1
\usepackage[usenames,dvipsnames]{color}  
\usepackage{graphicx}
\usepackage{caption}
\usepackage{subcaption}
\usepackage{amsmath}
\usepackage{amssymb} 
\usepackage{enumitem}
\usepackage[colorlinks=true,citecolor=darkgreen,urlcolor=darkgreen, pdfborder={0 0 0}]{hyperref}
\usepackage[normalem]{ulem}
\usepackage{hyperref}
\hypersetup{colorlinks=true,citecolor=green,linkcolor=blue,urlcolor=blue}
\usepackage{soul}
\usepackage{ulem}

\usepackage[dvipsnames]{xcolor}

\def\gsim{\raise0.3ex\hbox{$\;>$\kern-0.75em\raise-1.1ex\hbox{$\sim\;$}}}
\def\lsim{\raise0.3ex\hbox{$\;<$\kern-0.75em\raise-1.1ex\hbox{$\sim\;$}}}

\def\beqn#1{\begin{equation}\label{#1}}
\def\eeqn{\end{equation}}

\def\beqa#1{\begin{eqnarray}\label{#1}}
\def\eeqa{\end{eqnarray}}

\def\be{\begin{equation}}
\def\ee{\end{equation}}
\def\bea{\begin{eqnarray}}
\def\eea{\end{eqnarray}}

\def\Z2{$\mathcal{Z_2}$}

\newcommand {\ignore}[1]{}

\def\321{$\mathrm{SU(3) \otimes SU(2) \otimes U(1)}$ }

\def\tc#1{\textcolor{black}{#1}}
 
\begin{document}

\bibliographystyle{unsrt}   

\title{\boldmath Nonstandard neutrino interactions and mu-tau reflection symmetry}

\author{Jiajun Liao}
\email[Email: ]{liaojiajun@mail.sysu.edu.cn}
\affiliation{School of Physics,  Sun Yat-Sen University,  Guangzhou 510275,  China}
\author{Newton Nath}
\email[Email: ]{newton@ihep.ac.cn}
\affiliation{
Institute of High Energy Physics, Chinese Academy of Sciences, Beijing, 100049, China}
\affiliation{
School of Physical Sciences, University of Chinese Academy of Sciences, Beijing, 100049, China}
\author{TseChun Wang} 
\email[Email: ]{wangzejun@mail.sysu.edu.cn}
\affiliation{School of Physics,  Sun Yat-Sen University,  Guangzhou 510275,  China}
\author{Ye-Ling Zhou}
\email[Email: ]{ye-ling.zhou@soton.ac.uk}
\affiliation{School of Physics and Astronomy, University of Southampton, Southampton SO17 1BJ, United Kingdom}

\begin{abstract}

\vspace{1.0cm}
{\noindent
Nonstandard interactions (NSIs), possible subleading effects originating from new physics beyond the Standard Model, may affect the propagation of neutrinos and eventually contribute to measurements of neutrino oscillations. Besides this, $ \mu-\tau $ reflection symmetry, naturally predicted by non-Abelian discrete flavor symmetries, has been very successful in explaining the observed leptonic mixing patterns. In this work, we study the combined effect of both. We present an $S_4$ flavor model with $\mu-\tau$ reflection symmetry realized in both neutrino masses and NSIs. Under this formalism, we perform a detailed study for the upcoming neutrino experiments DUNE and T2HK. Our simulation results show that under the $\mu-\tau $ reflection symmetry, NSI parameters are further constrained and the mass ordering sensitivity is less affected by the presence of NSIs.}
\end{abstract}
\maketitle


\newpage

\section{Introduction}

 The observation of neutrino oscillations by various experiments using solar, atmospheric, reactor, and accelerator neutrinos 
 indubitably established the existence of neutrino mass, which
 provides a clear evidence of new physics beyond the Standard Model (SM)~\cite{Tanabashi:2018oca}. 
The effective Hamiltonian to describe neutrino oscillations is in general written as 
\begin{equation} \label{eq:Hamiltonian}
H = {1\over2E} \left( M_\nu M^{\dagger}_\nu + V\right)\,.
\end{equation}
Here, $M_\nu$ is the neutrino mass matrix and $V$ represents the effective potential for neutrino propagating in matter. Considering only the standard matter effect,  $V= {\rm diag}\{ A, 0 , 0 \}$ with $A = 2\sqrt2 G_F N_e E$, $N_e$ the electron number density in matter and $E$ the neutrino energy. 

Neutrino oscillation data have been inspiring theoretical studies of lepton mixing and mechanisms behind. Traditionally, the observation of maximal atmospheric mixing motivates the notion of the $\mu-\tau$ interchange symmetry, i.e., the permutation between the mu neutrino and the tau neutrino $\nu_\mu \leftrightarrow\nu_\tau$, which predicts $\theta_{23}= 45^\circ$ but $\theta_{13}=0$~\cite{Fukuyama:1997ky, *Ma:2001mr, *Lam:2001fb,*Balaji:2001ex,*Harrison:2002er}. 
Later, with more precise oscillation measurements, a critical nonvanishing reactor angle and a large CP violation were measured. The so-called the $\mu-\tau$ reflection symmetry, originally proposed in Ref.~\cite{Harrison:2002et} (for recent review, see ~\cite{Xing:2015fdg} and the references therein), has caught more attention.
This symmetry, on the phenomenological side, satisfies 
$ |U_{\mu i }| = |U_{\tau i }|$ (for $i = 1, 2, 3$), which predicts four cases~\cite{Liao:2012xm}: 
(a) \(\theta_{23}=45^\circ, \theta_{13}=0\); (b) \(\theta_{23}=45^\circ,
\theta_{12}=0\); (c) \(\theta_{23}=45^\circ, \theta_{12}=90^\circ\);
(d) \(\theta_{23}=45^\circ, \delta=\pm 90^\circ\).
The last case, specifically, $\theta_{23}^{} = 45^\circ, \delta = - 90^\circ$, is still in excellent agreement with the latest global analysis of neutrino oscillation data~\cite{Capozzi:2016rtj,*Esteban:2016qun,*deSalas:2017kay,*Esteban:2018azc}. 

Various non-Abelian discrete flavor models have been proposed to realize the $\mu-\tau$ reflection symmetry. Among them, the littlest $\mu-\tau$ seesaw model is the most predictive one, where all mixing parameters and the ratio $\Delta m^2_{21}/ \Delta m^2_{31}$ are dependent upon a single RG running parameter \cite{King:2018kka, King:2019tbt}. More models have been worked out in the framework of generalized CP symmetry \cite{Feruglio:2012cw,Holthausen:2012dk}. Combining the $\mu-\tau$ interchange with CP transformation, we arrive at the $\mu-\tau$ reflection transformation, 
\begin{eqnarray}   
\nu_e \to \nu_e^c \,,\quad
\nu_\mu \to \nu_\tau^c \,, \quad
\nu_\tau \to \nu_\mu^c \,.
\end{eqnarray}
Requiring neutrino mass matrix to be invariant under this transformation is equivalent to imposing the following constraint (assuming Majorana neutrinos):
\begin{equation}
X_{\mu\tau}^{T} M_{\nu} X_{\mu\tau} = M^*_{\nu} \,, 
\end{equation}
where 
\begin{equation}
X_{\mu\tau} = \begin{pmatrix}
1 & 0 & 0 \\
0  & 0 & 1 \\
0 & 1 & 0  \\
\end{pmatrix}\,.
\end{equation}
This immediately leads to
\begin{eqnarray}\label{eq:low_mnu}
M_\nu =  \left( \begin{matrix}
a &  b &  b^{*} \cr
b  & c  & d \cr  
b^{*} &  d & c^{*}  
\end{matrix} \right)  \;,
\end{eqnarray}
where $ a, d$ are real and $ b, c$ are complex. It predicts not just \(\theta_{23}=45^\circ, \delta=\pm 90^\circ\) with nonzero $\theta_{13}$, 
but also the Majorana phases $\rho, \sigma = 0, 90^\circ$ \cite{Zhou:2014sya}.
It is to be noted that a general strategy to test such symmetry has been examined in \cite{Srivastava:2018ser} for DUNE, where the predictions of $\mu-\tau$ reflection symmetry have arisen from discrete flavor group like $A_4$. On the other hand, the model-independent consequences of such symmetry have been pointed out in \cite{Nath:2018xkz} for DUNE. We notice from 
Refs.~\cite{Srivastava:2018ser,Nath:2018xkz} that depending on the representative values of $\theta_{23}$ and $\delta$,  DUNE can test such symmetry at more than 2$\sigma$ C.L.

Moreover, symmetry-based arguments on how to realize the mass texture $ M_\nu $, as given by Eq.~\eqref{eq:low_mnu}, have been discussed  in case of generalized CP combined with different non-Abelian discrete symmetries for $A_4$ \cite{Ding:2013bpa}, $S_4$ \cite{Holthausen:2012dk,Ding:2013hpa}, $\Delta(48)$ \cite{Ding:2013nsa, *Ding:2014hva}, and $\Delta(96)$ \cite{Ding:2014ssa}. Numerous examples of explicit model construction can be found in, e.g., \cite{Ding:2013bpa,Ding:2013hpa, Feruglio:2013hia, Li:2014eia}. 

Besides this, new physics beyond the SM may lead to corrections to the effective neutrino interactions through higher-dimensional operators.
These operators are often described in the framework of nonstandard interactions (NSIs). 
Here, we focus on NSI that can impact the propagation of neutrinos through matter. For neutrinos with kinetic energy around or below GeV scale, they are described by the dimension-6 four-fermion operators of the form \cite{Wolfenstein:1977ue}
\begin{equation}
\label{eq:NSI}
\mathcal
{L}_\text{NSI} = -2\sqrt{2} G_F 
(\overline{\nu}_\alpha \gamma^{\rho} 
P_L \nu_\beta)
(\bar{f} \gamma_{\rho} P_C f)
\epsilon^{fC}_{\alpha\beta} 
+ \text{H.c.}\,
\end{equation}
where $\epsilon^{f C}_{\alpha\beta}$ represent NSI parameters with 
$\alpha, \beta = e, \mu, \tau$, $C = L,R$, $f = e, u, d$, and 
$G_{F}$ is the Fermi constant. In the presence of  NSI, the effective potential $V$ in Eq.~\eqref{eq:Hamiltonian} is modified, i.e.,
\be
V =  A \left(\begin{array}{ccc}
	1 + \epsilon_{ee} & \epsilon_{e\mu} & \epsilon_{e\tau}
	\\
	\epsilon_{\mu e}& \epsilon_{\mu\mu} & \epsilon_{\mu\tau}
	\\
	\epsilon_{\tau e}& \epsilon_{\tau\mu} & \epsilon_{\tau\tau}
\end{array}\right)\,,
\label{eq:V_general}
\ee
where the $3\times 3$ NSI matrix $\epsilon_{\alpha\beta}$ with $\alpha, \beta = e,\mu,\tau$ is a Hermitian matrix, $\epsilon_{\alpha\beta} = \epsilon_{\beta\alpha}^*$.
The next-generation long baseline (LBL) accelerator neutrino oscillation experiments, DUNE~\cite{Acciarri:2015uup}  and T2HK~\cite{Abe:2018uyc}, aim to do precision tests for the standard three-flavor neutrino oscillation parameters. Moreover, because of their statistically high potential, they will also reach the sensitivity to probe NSIs. 

In recent times, there have been many interesting studies involving NSIs and their impacts on the measurement of standard neutrino oscillation parameters; for detailed reviews see 
Refs.~\cite{Ohlsson:2012kf,*Miranda:2015dra,*Farzan:2017xzy} and the references therein.
Authors of \cite{Liao:2016hsa,*Blennow:2016etl,*Agarwalla:2016fkh,*Masud:2016gcl,*Ghosh:2017ged,*Flores:2018kwk,*Masud:2018pig} have addressed various parameter degeneracies between standard and nonstandard interactions for the determination of neutrino mass ordering and the atmospheric mixing angle in case of DUNE and T2HK.
In particular, Ref.~\cite{Liao:2016orc} shows that when $|\epsilon_{e \mu}|=\tan\theta_{23}|\epsilon_{e \tau}|$, a cancellation between leading order terms in the appearance channel probabilities will strongly affect the sensitivities to NSI parameters at T2HK. 
Also, the generalized mass ordering degeneracy has been studied detailedly in~\cite{Bakhti:2014pva,*Coloma:2016gei,*Deepthi:2016erc}.
Besides this, some exhaustive analyses in the presence of  complex NSI parameters on the determination of the Dirac CP-violating phase have been performed in~\cite{Coloma:2015kiu,*Masud:2015xva,*deGouvea:2015ndi,*Huitu:2016bmb,*Forero:2016cmb,*Masud:2016bvp,*Ge:2016dlx,*Deepthi:2017gxg,*Capozzi:2019iqn}. Notice that the above-mentioned works are mostly based on model-independent studies, whereas some model-dependent analyses considering heavy charged singlet and/or doublet scalars have been performed in~\cite{Forero:2016ghr,*Dey:2018yht,*Bischer:2018zbd,*Babu:2019mfe}. In these studies, all NSI parameters are assumed to be free complex parameters. Such a large redundancy leads to a difficulty of making definite predictions. 

As pointed out in \cite{Wang:2018dwk}, the NSI effects may provide important information to extend our understanding of discrete flavor symmetries. A combined study of the flavor symmetry and NSIs can be pursued in the future neutrino experiments. 
As a result, once the $\mu-\tau$ reflection symmetry is a true symmetry in the neutrino mass matrix, one may wonder if NSIs also satisfy this symmetry. 
Furthermore, imposing the $\mu-\tau$ reflection symmetry on the NSI matrix can efficiently reduce the parameter redundancy and satisfy the condition $|\epsilon_{e \mu}|=\tan\theta_{23}|\epsilon_{e \tau}|$ given in~\cite{Liao:2016orc}. In this paper, we make an attempt to study the importance of NSIs for the upcoming LBL neutrino oscillation experiments DUNE and T2HK in the presence of  $\mu-\tau $ reflection symmetry. More specifically, we consider the NSI matrix that obeys $\mu-\tau $ reflection symmetry and within this framework, the NSI effect has been examined in the context of LBL experiments.

We outline the rest of this work  as follows. 
In Sec.~\ref{sec:FS}, for illustration, we show how to realize $\mu-\tau $ reflection symmetry in both neutrino mass and NSI matrixes in an $S_4$ flavor model. We follow the technique developed in Ref.~\cite{Wang:2018dwk}, which is the first paper in the literature which combines NSI with flavor symmetries together. 
Furthermore, within the $ S_4 $ flavor symmetry and the CP symmetry, we show that it is possible to realize NSI matrix that follows  $\mu-\tau $ reflection symmetry. 
In Sec.~\ref{sec:pheno}, we perform a phenomenological study of the model both analytically and numerically. Starting from the analysis of the oscillation probabilities, 
we study the constraints on $\epsilon_{\alpha\beta}$ and then 
measure
the impacts of NSI on the mass ordering sensitivity. 
We summarize our results in Sec.~\ref{sec:summary}.


\section{The mu-tau reflection symmetry driven by flavor symmetries\label{sec:FS}}

For illustration, we will construct a flavor model in which the $\mu-\tau$ reflection symmetry is preserved in both neutrino mass and NSI matrixes in the flavor basis. 
In order to generate the $\mu-\tau$ reflection symmetry in the neutrino mass term, we introduce the flavor symmetry $S_4 \times Z_4$. $S_4$ is the permutation group of four objects, whose generators $S$, $T$ and $U$ satisfying $T^3=S^2=U^2=(ST)^3=(SU)^2=(TU)^2=1$. Its properties are briefly listed in the Appendix.  We introduce necessary flavon fields which gain \tc{vacuum expectation values (VEVs)}, and as a consequence, the flavor symmetry is broken and a nontrivial flavor mixing is obtained. 
In the electroweak space, we imply the idea of scalar\tc{-and-}doublet-singlet mixing to generate sizable NSI\tc{s} \cite{Forero:2016ghr}, where an additional global symmetry, $Z_2$, is combined to avoid a myriad of experimental constraints from the charged lepton sector. 

We give particle contents of the model in Table~\ref{tab:particles}. Three lepton gauge doublets $L=(L_1,L_2,L_3)^T$ are arranged as a triplet $\mathbf{3}$ of $S_4$. Three right-handed charged leptons $e_R$, $\mu_R$ and $\tau_R$ are singlets of $S_4$. Four flavon fields are introduced: $\varphi$ is a triplet $\mathbf{3}'$ of $S_4$ to generate charged lepton Yukawa coupling; $\chi$, $\zeta$ and $\xi$ are triplet $\mathbf{3}$, doublet $\mathbf{2}$ and singlet $\mathbf{1}$, respectively, to generate the neutrino mass texture. The scalar\tc{-and-}doublet-singlet mixing is realized by introducing new electroweak doublets $\eta$ and a charged gauge singlet scalar $\phi^+$, respectively. Here, we arrange $\eta$ and $\phi^+$ as a triplet and a singlet of $S_4$, respectively. We discuss how to realize neutrino mass texture and the effective Hamiltonian terms of NSI with the $\mu-\tau$ reflection symmetry in the following two subsections, respectively. 
\begin{table}[h!]
\centering
\begin{tabular}{ |c|cccc|ccc|cccc|}\hline\hline
Fields & $L$ & $e_R$ & $\mu_R$ & $\tau_R$ & $H$ & $\eta$ & $\phi^+$ & $\varphi$ & $\chi$ & $\zeta$ & $\xi$ \\\hline

$SU(2)_L$ & $\mathbf{2}$ & $\mathbf{1}$ & $\mathbf{1}$ & $\mathbf{1}$ & $\mathbf{2}$ & $\mathbf{2}$ & $\mathbf{1}$ & $\mathbf{1}$ & $\mathbf{1}$ & $\mathbf{1}$ & $\mathbf{1}$ \\

$U(1)_Y$ & $-\frac{1}{2}$ & $-1$ & $-1$ & $-1$ & $+\frac{1}{2}$ & $+\frac{1}{2}$ & $+1$ & $0$ & $0$ & $0$ & $0$ \\\hline

$S_4$ & $\mathbf{3}$ & $\mathbf{1}$ & $\mathbf{1}$ & $\mathbf{1}'$ & $\mathbf{1}$ & $\mathbf{3}$ & $\mathbf{1}$ & $\mathbf{3}'$ & $\mathbf{3}$ & $\mathbf{2}$ & $\mathbf{1}$ \\

$Z_4$ & $1$ & $i$ & $-1$ & $-i$ & $1$ & $-i$ & $-i$ & $i$ & $1$ & $1$ & $1$ \\\hline

$Z_2$ & $+$ & $-$ & $+$ & $-$ & $+$ & $-$ & $-$ & $-$ & $+$ & $+$ & $+$ \\\hline\hline
\end{tabular} 
  \caption{\label{tab:particles} Particles contents and their assignments in the electroweak gauge symmetry $SU(2)_L \times U(1)_Y$ and flavor symmetry $S_4\times Z_4$ and an additional global symmetry $Z_2$. }
\end{table}


\subsection{The mu-tau reflection symmetry realized in neutrino mass terms}

We first consider the construction of a simplified neutrino mass model. To be invariant in $S_4$,  
the effective Lagrangian terms to generate charged lepton masses and neutrino masses should take the form (c.f. Eq.~\eqref{eq:mass_term} of the Appendix)
\begin{eqnarray} 
   &&-\mathcal{L} \supset
    \overline{L} H \left[
    \frac{y_e}{\Lambda^3} \!
    \begin{pmatrix} \varphi_3^3 -\varphi_2^3 \\ 
    2\varphi_1^2\varphi_2-\varphi_2^2\varphi_3-\varphi_3^2\varphi_1 \\ 
    -2\varphi_1^2\varphi_3+\varphi_2^2\varphi_1+\varphi_3^2\varphi_2 
    \end{pmatrix} 
     e_R + 
     \frac{y_\mu}{\Lambda^2} \!
     \begin{pmatrix} \varphi_1^2-\varphi_2\varphi_3 \\ \varphi_3^2-\varphi_1\varphi_2 \\ \varphi_2^2-\varphi_1\varphi_3 \end{pmatrix} 
     \mu_R + 
     \frac{y_\tau}{\Lambda} \!
     \begin{pmatrix} \varphi_1 \\ \varphi_2 \\ \varphi_3 \end{pmatrix} \tau_R \right] \nonumber\\
    &&+ \frac{1}{2\Lambda_{\rm ss}}\overline{L} \tilde{H} 
    \left[\frac{y_1}{\Lambda} \begin{pmatrix} \xi & 0 & 0 \\ 0 & 0 & \xi \\ 0 & \xi & 0 \end{pmatrix} + 
    \frac{y_2}{\Lambda} \begin{pmatrix} 0 & \zeta_1 & \zeta_2 \\ \zeta_1 & \zeta_2 & 0 \\ \zeta_2 & 0 & \zeta_1 \end{pmatrix} + 
    \frac{y_3}{\Lambda} \begin{pmatrix} 2\chi_1 & -\chi_2 & -\chi_3 \\ -\chi_2 & 2\chi_3 & -\chi_1 \\ -\chi_3 & -\chi_1 & 2\chi_2 \end{pmatrix}
    \right]
    H^* L^c
     + {\rm H.c.} \,, 
\end{eqnarray}
where $\overline{L} = \overline{(L_1,L_2,L_3)}$, $\tilde{H} = i\sigma_2 H^*$,
$\Lambda$ is the scale of heavy flavor multiplets decouple and $\Lambda_{\rm ss}$ is the seesaw scale where heavy mediators (e.g., right-handed neutrinos) decouple. 
This arrangement can be achieved by arranging additional $Z_n$ charges for flavons \cite{Ding:2013hpa,Feruglio:2013hia}. 
Combining the CP symmetry with the flavor symmetry together and keeping in mind that particles are flavor multiplets in the flavor space, all coefficients are forced to be real \cite{Feruglio:2012cw,Holthausen:2012dk}. 

Charged lepton masses are obtained after the flavon $\varphi$ and the Higgs gain VEVs. Given the VEV for the flavon in the charged lepton sector,
\begin{eqnarray}\label{eq:varphi_VEV}
    \langle \varphi_1 \rangle =
    \langle \varphi_2 \rangle= 0\,, ~ \langle \varphi_3 \rangle = v_\varphi \,,
\end{eqnarray}
a diagonal charged lepton mass matrix is obtained with diagonal entries given by 
\begin{eqnarray}
    m_e = \frac{y_e v_\varphi^3} {(2\sqrt{2}G_F)^{1/2}\Lambda^3} \,,~
    m_\mu = \frac{y_\mu v_\varphi^2} {(2\sqrt{2}G_F)^{1/2}\Lambda^2}\,,~
    m_\tau = \frac{y_\tau v_\varphi} {(2\sqrt{2}G_F)^{1/2} \Lambda}\,,
\end{eqnarray}
where the Higgs VEV $\langle H \rangle = v_H \approx  (2\sqrt{2}G_F)^{-1/2} =174$ GeV has been used.
In this case, the three lepton gauge doublets $L_1$, $L_2$ and $L_3$ are identified with the three-flavor states $L_e = (\nu_e,e)_L^T$, $L_\mu = (\nu_\mu,\mu)_L^T$, $L_\tau = (\nu_\tau,\tau)_L^T$, respectively.

Neutrinos obtain their masses after the flavons $\chi$, $\zeta$, $\xi$ and the Higgs $H$ gain VEVs. Since we assume that the $\mu-\tau$ reflection symmetry is a residual symmetry in the neutrino sector, VEVs of $\xi$, $\zeta$ and $\chi$ should be invariant under the relevant transformation, namely, 
\begin{eqnarray}\label{eq:vev_flavon}
    \langle \xi \rangle^* = \langle \xi \rangle \,,~~~
    X_{\mu\tau,\mathbf{2}} \begin{pmatrix} \langle \zeta_1 \rangle \\ \langle \zeta_2 \rangle \end{pmatrix}^*
    = 
    \begin{pmatrix} \langle \zeta_1 \rangle \\ \langle \zeta_2 \rangle \end{pmatrix} \,,~~~
    X_{\mu\tau} \begin{pmatrix} \langle \chi_1 \rangle \\ \langle \chi_2 \rangle \\ \langle \chi_3 \rangle \end{pmatrix}^* = 
    \begin{pmatrix} \langle \chi_1 \rangle \\ \langle \chi_2 \rangle \\ \langle \chi_3 \rangle \end{pmatrix}\,.
\end{eqnarray}
Here, $X_{\mu\tau,\mathbf{2}}$ and $X_{\mu\tau}$ are the representation-dependent transformation matrices and respect the $\mu-\tau$ reflection symmetry. They are correlated with the generator $U$, which represents the $\mu-\tau$ transformation. Given the representation matrices in Table~\ref{tab:rep_matrix2}, the two-dimensional $X_{\mu\tau,\mathbf{2}}$ and the three-dimensional $X_{\mu\tau}$ are given by \cite{Ding:2013hpa, Feruglio:2013hia}
\begin{eqnarray}
    X_{\mu\tau,\mathbf{2}} = \begin{pmatrix} 0 & 1 \\ 1 & 0 \end{pmatrix} \,,~~
    X_{\mu\tau} = \begin{pmatrix} 1 & 0 & 0 \\ 0 & 0 & 1 \\ 0 & 1 & 0 \end{pmatrix} \,.
\end{eqnarray}
The condition in Eq.~\eqref{eq:vev_flavon} leads to real $\langle \xi \rangle$, $\langle \chi_1 \rangle$ and complex-conjugate pairs $\langle \zeta_1 \rangle^* = \langle \zeta_2 \rangle$, $\langle \chi_2 \rangle^*=\langle \chi_3 \rangle$. We denote them as 
\begin{eqnarray}\label{eq:VEVs}
    &&\langle \xi \rangle = v_1\,,\hspace{14mm} \langle \chi_1 \rangle = u_1 \,, \nonumber\\
    &&\langle \zeta_1 \rangle =  v_2 + i v_3\,,~~\, \langle \zeta_2 \rangle = v_2 - i v_3\,,~~\nonumber\\
    &&
    \langle \chi_2 \rangle = u_2 + i u_3 \,, ~
    \langle \chi_3 \rangle = u_2 - i u_3 \,,
\end{eqnarray}
where $v_i$ and $u_i$ (for $i=1,2,3$) are all real. 

The special directions of these VEVs can be justified via $S_4$-invariant terms in the flavon potential. For example, we consider the following cubic and quartic terms in the potential to fix directions of $\langle \zeta \rangle$ and $\langle \chi \rangle$:
	\begin{eqnarray} \label{eq:potential}
	\hspace{-5mm} 
	V \supset &&\lambda_1 [(\tilde{\zeta} \zeta)_{\mathbf{1}'}]^2 
	+ \lambda_2 \Big[ \big( (\zeta \zeta)_{\mathbf{2}}(\tilde{\zeta} \zeta)_{\mathbf{2}}\big)_{\mathbf{1}} + {\rm h.c.}\Big]
	+ \lambda_3 \big((\tilde{\chi} \chi)_{\mathbf{3}'} (\tilde{\chi} \chi)_{\mathbf{3}'}\big)_{\mathbf{1}} \nonumber\\
	&&+ \lambda_4 \Big[ \big(\chi (\tilde{\chi} \chi)_{\mathbf{3}})_{\mathbf{3}}\big)_{\mathbf{1}} + {\rm h.c.}\Big]\,,
	\end{eqnarray}
	with $\lambda_1, \lambda_3>0$, and $\lambda_2<0$ satisfied.  Here, the subscripts represent contractions to irreducible representations of $S_4$, and we have written conjugates of $\zeta = (\zeta_1, \zeta_2)^T$ and $\chi = (\chi_1, \chi_2, \chi_3)^T$ in the representation basis of $S_4$: $\tilde{\zeta}=(\zeta_2^*, \zeta_1^*)$ and $\tilde{\chi} = (\chi_1^*, \chi_3^*, \chi_2^*)^T$. See the Appendix~\ref{app:S4} for more details. The first term in Eq.~\eqref{eq:potential} is identical to $\lambda_1 (|\zeta_1|^2 - |\zeta_2|^2)^2$, which is minimized at $|\zeta_1| =  |\zeta_2|$. Taken this into account, the second term is simplified to $4 \lambda_2 |\zeta_1|^4 \cos (\alpha_1+\alpha_2)$, where $\alpha_1$ and $\alpha_2$ are phases of $\zeta_1$ and $\zeta_2$, respectively. It  is minimal at $\alpha_1 = - \alpha_2$, leading to $\langle \zeta_1 \rangle = \langle \zeta_2 \rangle^*$. The VEV $\langle \chi \rangle$ could be partially determined by the third term of Eq.~\eqref{eq:potential}, which is expanded into
	\begin{eqnarray} \label{eq:lambda_3}
	\lambda_3 \left[ (|\chi_2|^2 - |\chi_3|^2)^2 + 2 |\chi_1|^2 (|\chi_2|^2 + |\chi_3|^2)
	- 4 |\chi_1^2 \chi_2 \chi_3| \cos (2\beta_1 - \beta_2 - \beta_3) \right]\,,
	\end{eqnarray}
	where $\beta_1$, $\beta_2$, and $\beta_3$ are phases of $\chi_1$, $\chi_2$, and $\chi_3$, respectively. The first term in Eq.~\eqref{eq:lambda_3} takes the minimal value at $|\chi_2| = |\chi_3|$. After fixing the absolute values of $\chi_1$ and $\chi_2$, we find that the third term is minimized at $2\beta_1 - \beta_2 - \beta_3 = 0$.
	We further consider the fourth term of Eq.~\eqref{eq:potential}, now simplified to be $4\lambda_4 \cos\beta_1\, [|\chi_1|^3- 3 |\chi_1 \chi_2^2| \cos(3\beta_1 - 3\beta_2)]$. Depending on absolute values of $\chi_1$ and $\chi_2$ and the sign of $\lambda_4$. This term can be either maximum or minimum at $\beta_1=0$. Once $\beta_1=0$ is fixed, we arrive at $\beta_2 = - \beta_3$. Therefore, we realize the VEV of $\chi$.

After flavons gain VEVs in Eq.~\eqref{eq:VEVs}, we arrive at the  neutrino mass matrix in Eq.~\eqref{eq:low_mnu} with  
\begin{eqnarray}
&& a = \frac{y_1 v_1+ 2 y_3 u_1}{2\sqrt{2}G_F \Lambda_{\rm ss} \Lambda} \,,\quad
 b = \frac{y_2 v_2 - y_3 u_2 + i (y_2 v_3 - y_3 u_3)}{2\sqrt{2}G_F \Lambda_{\rm ss} \Lambda} \,, \nonumber\\
&& c = \frac{y_2 v_2 + 2 y_3 u_2 - i (y_2 v_3 + 2y_3 u_3)}{2\sqrt{2}G_F \Lambda_{\rm ss} \Lambda} \,,\quad
 d = \frac{y_1 v_1 - y_3 u_1}{2\sqrt{2}G_F \Lambda_{\rm ss} \Lambda} \,.
\end{eqnarray} 
This is the most general neutrino mass matrix preserving the $\mu-\tau$ reflection symmetry.
It is straightforward to check  its invariance under the $\mu-\tau$ reflection transformation \cite{Zhou:2014sya},
\begin{eqnarray}
    \nu_e \to \nu_e^c \,,~ 
    \nu_\mu \to \nu_\tau^c \,,~ 
    \nu_\tau \to \nu_\mu^c \,.
\end{eqnarray}
It is equivalent to require 
\begin{eqnarray}
    X_{\mu\tau}^T M_\nu X_{\mu\tau} = M_\nu^* \,.
\end{eqnarray}

This simplified model shows a generic way to achieve flavor mixing with the $\mu-\tau$ reflection symmetry. 
For an explicit flavor model construction, one should carefully consider how to achieve the required flavon VEVs and how to avoid the unnecessary higher-dimensional operators. In order to resolve these issues, supersymmetry and extra fields have to be included, extra $Z_n$ symmetries may be imposed, and \tc{an} UV completion of the effective theory may be considered. For relevant discussions, see e.g., \cite{Ding:2013hpa,Feruglio:2013hia, Li:2014eia}. As a consequence, correlations between oscillation parameters are induced in these models, which can be tested in the future neutrino oscillation experiments. In this paper, we will not consider these parameter correlations. 

\subsection{The mu-tau reflection symmetry realized in nonstandard interactions \label{sec:mutau_NSI}}

We further discuss how to achieve NSIs satisfying the $\mu-\tau$ reflection symmetry. We couple flavons to the NSI mediator charged scalars $\eta$ and $\phi^+$. Since VEVs of flavons contributing to neutrino masses satisfy the $\mu-\tau$ reflection symmetry, any interactions between these flavons and charged scalars satisfy the $\mu-\tau$ reflection symmetry. 

In order to see the behavior of the $\mu-\tau$ reflection symmetry in NSIs, we first consider the following benchmark Lagrangian terms for illustration\footnote{In the basis as shown in the Appendix, the conjugate of $L$ should be arranged as $\overline{L} = (\overline{L}_e, \overline{L}_\tau, \overline{L}_\mu)^T$ \cite{deMedeirosVarzielas:2017hen,Wang:2018dwk}.}:
\begin{eqnarray} \label{eq:charged_scalar_Lagrangian}
    -\mathcal{L} &\supset& \mu_\eta^2 \eta^\dag \eta + \mu_\phi^2 \phi^- \phi^+ 
    +\left( f \phi^- \tilde{H}^\dag (\eta_1 \chi_1 + \eta_2 \chi_3 + \eta_3 \chi_2)  + {\rm H.c.}\right) \nonumber\\
    &&+\lambda (\overline{L}_e \eta_1 + \overline{L}_\mu \eta_2 + \overline{L}_\tau \eta_3)
    e_R \,,
\end{eqnarray}
where all coefficients are real by imposing CP symmetry. 
The first row gives rise to charged scalar masses after the SM Higgs obtain the VEV. In the basis $S^+ = (\eta_1^+, \eta_2^+, \eta_3^+, \phi^+)^T$, the charged scalar mass matrix is given by
\begin{eqnarray}\label{eq:MS_matrix}
    M_{S+}^2 = \begin{pmatrix} 
    \mu_\eta^2 & 0 & 0 & f u_1 v_H\\ 
    0 & \mu_\eta^2 & 0 & f (u_2 + i u_3) v_H\\
    0 & 0 & \mu_\eta^2 & f (u_2 - i u_3) v_H\\
    f u_1 v_H & f (u_2 - i u_3) v_H & f (u_2 + i u_3) v_H & \mu_\phi^2 \end{pmatrix} \,.
\end{eqnarray}
Here, the $Z_2$ symmetry is not broken after flavor symmetry breaking and electroweak symmetry breaking. 
The second row of Eq.~\eqref{eq:charged_scalar_Lagrangian} gives rise to the interaction of charged scalars, neutrinos and the electron. It is explicitly written out as
\begin{eqnarray}
    -\mathcal{L} \supset \overline{(\nu_e, \nu_\mu, \nu_\tau)_L} P_\lambda S^+ e_R,
\end{eqnarray}
where the coefficient matrix $P_\lambda$ is given by
\begin{eqnarray}
    P_\lambda = \lambda \begin{pmatrix} 1 & 0 & 0 & 0 \\ 0 & 1 & 0 & 0 \\ 0 & 0 & 1 & 0 \end{pmatrix}\,.
\end{eqnarray}
The charged scalars have masses $\gtrsim \mathcal{O}(100)$ GeV. For GeV-scale neutrino beams propagating in the Earth, these scalars can be integrated and dimension-six operators are left, 
\begin{eqnarray}
    \mathcal{L}_{\rm NSI} = \overline{(\nu_e, \nu_\mu, \mu_\tau)_L} e_R \, 
    \left[ P_\lambda ( M_{S^+}^2 )^{-1} P_\lambda^\dag \right] \overline{e_R}\begin{pmatrix}\nu_e \\ \nu_\mu \\ \mu_\tau \end{pmatrix}_L \,.
\end{eqnarray}
With the help of Fierz identity, we express it in the form 
\begin{eqnarray}
    \mathcal{L}_{\rm NSI}= - 2 \sqrt{2} G_F \sum_{\alpha,\beta} \epsilon_{\alpha\beta}^{eR}\, (\overline{\nu_{\alpha}} \gamma_\mu P_L \nu_{\beta}) ~  (\overline{e} \gamma^\mu P_R e) \,,
\label{eq:epsR}
\end{eqnarray}
where
\begin{eqnarray}\label{eq:nsimatrix}
    \epsilon^{e R} = \frac{-1}{4\sqrt{2} G_F} P_\lambda ( M_{S^+}^2 )^{-1} P_\lambda^\dag
    =
    \begin{pmatrix}
    \epsilon_{ee} & \epsilon_{e\mu} & \epsilon_{e\mu}^* \\
    \epsilon_{e\mu}^* & \epsilon_{\mu\mu} & \epsilon_{\mu\tau} \\
    \epsilon_{e\mu} & \epsilon_{\mu\tau}^* & \epsilon_{\mu\mu}
    \end{pmatrix},
\end{eqnarray}
with
\begin{eqnarray}\label{eq:nsiHermitian}
    &&\epsilon_{ee} = 2 a f^2 \left(u_2^2+u_3^2\right) - 2\sqrt{2} a G_F \mu_\eta ^2 \mu_\phi ^2
    \,,    \quad
    \epsilon_{e\mu} = - a f^2 u_1 (u_2-i u_3) \,,\nonumber\\
    &&\epsilon_{\mu\mu} = a f^2 \left(u_1^2+u_2^2+u_3^2\right) - 2\sqrt{2} a G_F \mu_\eta ^2 \mu_\phi ^2
    \,, \quad
    \epsilon_{\mu\tau} = - a f^2 (u_2+i u_3)^2 \,,
\end{eqnarray}
where $a^{-1} = 4\sqrt{2} G_F \mu_\eta^2 [f^2(u_1^2 + 2 u_2^2 + 2 u_3^2) - 2\sqrt{2} G_F \mu_\eta^2 \mu_\phi^2]$, and the Higgs VEV again has been replaced with $(2\sqrt{2}G_F)^{-1/2}$. In this benchmark, the $\mu-\tau$ reflection symmetry  in NSIs originates from  the contribution of $\chi$ to the off-diagonal mass term between $\eta$ and $\phi^+$. A correlation between NSI parameters $2{\rm Arg}(\epsilon_{e\mu}) + {\rm Arg} (\epsilon_{\mu\tau}) = 0$ holds, since only one single source of the $\mu-\tau$ reflection symmetry, i.e., the single coupling with $\chi$,  is included. This correlation can be removed easily by including additional interactions between flavons ($\xi$, $\zeta$, and $\chi$) and charged scalars. Assuming no interactions between the flavon $\varphi$ and electric-charged scalars, which can be achieved by assuming $Z_n$ charges in the supersymmetry framework, UV completions of NSIs do not induce any source to break the $\mu-\tau$ reflection symmetry, and eventually the $\mu-\tau$ reflection symmetry is always preserved. 

This model implies the advantage of using the scalar mixing between electroweak doublets and charged singlets to realize relatively sizable NSI, $\epsilon_{\alpha\beta} \sim \mathcal{O}(0.1)$. As originally proposed in \cite{Forero:2016ghr}, sizable NSIs require a price of fine-tuning between a large mixing and hierarchical masses for scalars, e.g., the mixing angle around $0.3$, the light charged scalar mass around 100 GeV, and the heavier charged scalar mass around 10 TeV. Here the light scalar mass is consistent with the lowest bound ($80$~GeV) of charged Higgs searches at LEP \cite{Abbiendi:2013hk}, and the heavier scalar mass satisfies constraint coming from various intensity experiments discussed below. Comparing with the original model in \cite{Forero:2016ghr}, we extend the gauge doublet $\eta$ from a flavor singlet to a flavor triplet of $S_4$. We comment that at the price of fine-tuning, NSI $\epsilon_{\alpha\beta} \sim \mathcal{O}(0.1)$ can still be achieved. While the constraints to the light scalar mass keeps the same (lowest bound at $80$~GeV),we will analyze in detail how these intensity experiments constrain heavy scalar masses.

Constraints to heavy neutral scalars are obtained from the precision measurement of $e^+ e^-\to \ell^+ \ell^-$ scatterings and upper bounds of charged lepton flavor violation (LFV) decay widths. These channels are mediated by heavy neutral components $\eta^0 = (\eta_1^0, \eta_2^0, \eta_3^0)^T$, whose masses are approximate to heavy charged scalar masses in the limit $f u_{1,2,3} v_H \ll \mu_\eta^2$. 
The coupling $\lambda (\overline{L}_e \eta_1^0 + \overline{L}_\mu \eta_2^0 + \overline{L}_\tau \eta_3^0)e_R$ is the source of these processes. Integrating out the heavy neutral scalars, effective four-charged-lepton interactions  are written as
\begin{eqnarray}
\frac{\lambda^2}{2}
\overline{(e,~ \mu,~ \tau)} \, 
    (M_{\eta^0}^2)^{-1} \gamma_\mu P_L \begin{pmatrix}e \\ \mu \\ \tau \end{pmatrix} \overline{e} \gamma^\mu P_R e \,,
\end{eqnarray}
where $M_{\eta^0}^2$ is the $3\times 3$ mass matrix of $\eta^0$. 
A diagonal $M_{\eta^0}^2$ induces only corrections to $e^+ e^-\to \ell^+ \ell^-$, but no charged LFV decays at tree level. 
New physics scales for  $e^+ e^-\to e^+ e^-$ and $e^+ e^-\to \mu^+ \mu^-, \tau^+ \tau^-$ have been pushed up to 9.1 and 10.2~TeV, respectively \cite{LEP:2003aa}, which transform to the constraints \cite{Forero:2016ghr}
\begin{equation}
\lambda^2[(M_{\eta^0}^2)^{-1}]_{ee} < (0.39/{\rm TeV})^2\,,\qquad 
\lambda^2[(M_{\eta^0}^2)^{-1}]_{\mu\mu}\,,~
\lambda^2[(M_{\eta^0}^2)^{-1}]_{\tau\tau} < (0.49/{\rm TeV})^2. 
\end{equation}
Here, $[(M_{\eta^0}^2)^{-1}]_{ee}$ is understood as the (1,1) entry of $(M_{\eta^0}^2)^{-1}$. 
In particular, our benchmark model in Eq.~\eqref{eq:charged_scalar_Lagrangian} gives $M_{\eta^0}^2 = \mu_\eta^2 {\rm diag}\{1,1,1\}$ and the scattering constraint $\lambda/\mu_\eta \lesssim 0.39/{\rm TeV}$. Note that the three heavy charged scalar masses are approximate to $\mu_\eta$. 
Including additional interactions between flavons and $\eta$ may generate off-diagonal entries of $M_{\eta^0}^2$ and induce $\tau^- \to e^-e^+e^-$, $\mu^- e^+e^-$ and $\mu^-\to e^-e^+e^-$ decays. 
Given the experimental constraints \cite{Hayasaka:2010np, Bellgardt:1987du}
\begin{eqnarray} 
&&{\rm BR}(\tau^- \to e^-e^+e^-) < 2.7 \times 10^{-8}\,, \nonumber\\
&&{\rm BR}(\tau^- \to\mu^- e^+e^-)< 1.8 \times 10^{-8}\,,\nonumber\\
&&{\rm BR}(\mu^-\to e^-e^+e^-)< 10^{-12}\,,
\end{eqnarray}
we obtain $\lambda^2[(M_{\eta^0}^2)^{-1}]_{e\tau} < (0.16/{\rm TeV})^2$, $\lambda^2[(M_{\eta^0}^2)^{-1}]_{\mu\tau} < (0.15/{\rm TeV})^2$, and $\lambda^2[(M_{\eta^0}^2)^{-1}]_{e\mu} < (0.013/{\rm TeV})^2$. 
Given a model with $\lambda \sim 1$ and $\mu_\eta \sim 10$~TeV, i.e., all neutral scalar and three charged scalar masses around 10 TeV, all the above constraints can be satisfied. 
We further comment that nonzero off-diagonal entries of $M_{\eta^0}^2$ also induce $\mu$ and $\tau$ radiative decays. All these channels are strongly suppressed by the electron mass involved in the loop and thus negligible.

\section{Testing Mu-tau reflection symmetry with NSI\label{sec:pheno}}
Here we study the phenomenological consequences of the $\mu-\tau $ reflection symmetry preserved in both the neutrino mass matrix and the NSI matrix using the upcoming neutrino oscillation experiments\textemdash DUNE and T2HK.
We start our phenomenological study from the analysis on oscillation probabilities with NSI under the $\mu-\tau$ reflection symmetry. In the following, we briefly introduce the experimental setup for experiments, and the analysis details, e.g., the true values and priors for oscillation parameters and $\epsilon_{\alpha\beta}$. Before viewing the impact of this symmetry on the mass ordering sensitivity, we investigate the restriction of the $\mu-\tau$ reflection symmetry on the allowed region of $\epsilon_{\alpha\beta}$.

\subsection{Neutrino oscillation probabilities}\label{sec:nsi}

In the framework of the $ \mu-\tau $ reflection symmetry, $M_\nu$ takes Eq.~\eqref{eq:low_mnu} and $\epsilon_{\alpha\beta}$ satisfy the correlations 
\begin{eqnarray}
\epsilon_{e\mu}^* = \epsilon_{e\tau} \,,\quad
\epsilon_{\mu\mu} = \epsilon_{\tau\tau} \,.
\end{eqnarray}
Matrices $M_\nu M^{\dagger}_\nu$ and $V$, appearing in the Hamiltonian, are expressed in forms of
\begin{eqnarray}\label{eq:herm_mnu}
M_\nu M^{\dagger}_\nu  =  \left( \begin{matrix}
\mathbf{a} & \mathbf{b}~ &  \mathbf{b}^{*} \cr
\mathbf{b}^*  & \mathbf{c}~  & \mathbf{d}~\cr  
\mathbf{b}~ & \mathbf{d}^* & \mathbf{c}~  
\end{matrix} \right)  \;, \quad
V =  A \left(\begin{array}{ccc}
	1 + \tilde{\epsilon}_{ee} & \epsilon_{e\mu} & \epsilon_{e\mu}^*
	\\
	\epsilon_{e\mu}^*& 0 & \epsilon_{\mu\tau}
	\\
	\epsilon_{e\mu}& \epsilon_{\mu\tau}^*& 0
\end{array}\right)\,,
\label{eq:V}
\end{eqnarray}
with $ \mathbf{a}$, $\mathbf{c}$ are real and $ \mathbf{b}$, $\mathbf{d}$ are complex. We have redefined $V$ with $\tilde{\epsilon}_{ee}\equiv\epsilon_{ee}-\epsilon_{\tau\tau} = \epsilon_{ee}-\epsilon_{\mu\mu}$ to absorb an overall diagonal phase. 
Note that in this symmetry-based formalism $V$ contains five real free NSI parameters, namely, one real ($ \tilde{\epsilon}_{ee} $) and two complex ($ \epsilon_{e\mu}, \epsilon_{\mu \tau} $) entries.
Bounds on these NSI parameters from the global analysis of oscillation data in
	solar, atmospheric, reactor, and long-baseline accelerator neutrino experiments have been performed in Ref.~\cite{Esteban:2018ppq}. Assuming Gaussian distributions, and taking the bounds from the 2$\sigma$ allowed ranges for $\epsilon_{\alpha\beta}^p$, we obtain the 90\% C.L. bounds on the NSI parameters as shown in Table~\ref{tab:bounds}. Note that since only the couplings between neutrinos and right-handed electrons are relevant in our model [c.f. Eq.~\eqref{eq:epsR}], the COHERENT~\cite{Akimov:2017ade} constraints in Ref.~\cite{Esteban:2018ppq}  do not apply here. For the same reason, we consider only the neutrino-electron scattering constraints from the LSND~\cite{Auerbach:2001wg} and the CHARM II~\cite{Vilain:1994qy} experiment. The 90\% C.L. bounds on the NSI parameters from Ref.~\cite{Davidson:2003ha} are shown in Table~\ref{tab:bounds}.\footnote{
		As discussed in Sec.~\ref{sec:mutau_NSI}, these NSI bounds can be saturated within our model at the cost of fine-tuning.}

%
	\begin{table}[]
		\begin{tabular}{|c|c|c|}
			\hline
			Parameters$~~~~~~~~~$ & Neutrino oscillation~~~~~ &Neutrino-electron scattering\\\hline
			$\tilde{\epsilon}_{ee}$$~~~~~~~~~$ & $[-0.034,~1.078]$$~~~~~$ & $[-1,0.5]$\\\hline
			$|\epsilon_{e\mu}|$$~~~~~~~~~$ & $[0,~0.146]$$~~~~~$ & $[0,~0.1]$\\\hline
			$|\epsilon_{e\tau}|$$~~~~~~~~~$ & $[0,~0.784]$$~~~~~$ & $[0,~0.7]$\\\hline
			$|\epsilon_{\mu\tau}|$$~~~~~~~~~$ & $[0,~0.029]$$~~~~~$ & $[0,~0.1]$\\\hline
		\end{tabular}
		\caption{90\% C.L. bounds on the NSI parameters adopted from the current global fit of neutrino oscillation data~\cite{Esteban:2018ppq} and neutrino-electron scattering experiments~\cite{Davidson:2003ha}.
		}
	\label{tab:bounds}
	\end{table}

The appearance probability for long baseline neutrino oscillation experiments in the presence of NSI is complicated in general~\cite{Liao:2016orc}. However, under the $\mu-\tau$ reflection symmetry, we can write the appearance probability for the normal ordering (NO) in a simple form, i.e.,
\bea
P_{\mu e} &= &x^2 f^2 + 2xyfg \cos(\Delta + \delta) + y^2 g^2
\nonumber\\
&-& 4\hat A fg|\epsilon_{e\mu}|
\left[ x  \sin\phi_{e\mu} \sin(\Delta+\delta) \right]
+ 4\hat A^2    f^2  |\epsilon_{e\mu}|^2 
- 4 \hat A^2 fg |\epsilon_{e\mu}|^2 \sin(2\phi_{e\mu})\sin\Delta
\nonumber\\
&+&   
{\cal O}(s_{13}^2 \epsilon, r\epsilon, s_{13}\epsilon^2, \epsilon^3)\,,
\label{eq:app-prob}
\eea
where $\phi_{\alpha\beta}\equiv\text{Arg}(\epsilon_{\alpha\beta})$ and following Ref.~\cite{Barger:2001yr},
\bea
x &\equiv& \sqrt{2} c_{13} s_{13}\,,\quad
y \equiv \sqrt{2}r s_{12} c_{12} \,,
\quad r = |\delta m^2_{21}/\delta m^2_{31}|\,,
\nonumber\\
f,\, \bar{f} &\equiv& \frac{\sin[\Delta(1\mp\hat A(1+\tilde{\epsilon}_{ee}))]}{(1\mp\hat A(1+\tilde{\epsilon}_{ee}))}\,,\ 
g \equiv \frac{\sin(\hat A(1+\tilde{\epsilon}_{ee}) \Delta)}{\hat A(1+\tilde{\epsilon}_{ee})}\,,\nonumber\\
\Delta &\equiv &\left|\frac{\delta m^2_{31} L}{4E}\right|,\ 
\hat A \equiv \left|\frac{A}{\delta m^2_{31}}\right|\,.
\label{eq:define}
\eea
The antineutrino probability $P_{\overline{\mu}^{}\overline{e}}$, which is given by
Eq.~(\ref{eq:app-prob}) with $\delta \to - \delta$, $\phi_{\alpha\beta} \to -
\phi_{\alpha\beta}$, and $\hat A \to - \hat A$ (and hence $f \to
\bar{f}$). For the inverted ordering (IO), $\Delta \to -
\Delta$, $y \to -y$, $\hat A \to - \hat A$ (and hence $f \leftrightarrow -
\bar{f}$, and $g \to -g$). 

For small $\epsilon$, we can neglect the $y^2$, $\epsilon^2$ and higher order terms in
Eq.~(\ref{eq:app-prob}). For $\delta=\pm 90^\circ$ and the NO, we have
\bea
(P_{\mu e})_\text{NO}&\simeq&x^2f^2\mp2xfg(y\sin\Delta+2\hat A |\epsilon_{e\mu}| \sin\phi_{e\mu}\cos\Delta)\,, \nonumber\\
(P_{\overline{\mu}^{}\overline{e}})_\text{NO}&\simeq&x^2\bar{f}^2\pm2x\bar{f}g(y\sin\Delta+2\hat A |\epsilon_{e\mu}| \sin\phi_{e\mu}\cos\Delta)\,.
\label{eq:PNO}
\eea
For $\delta=\pm 90^\circ$ and the IO, we have 
\bea
(P_{\mu e})_\text{IO}&\simeq&x^2\bar{f}^2\mp2x\bar{f}g(y\sin\Delta-2\hat A |\epsilon_{e\mu}| \sin\phi_{e\mu}\cos\Delta)\,, \nonumber\\
(P_{\overline{\mu}^{}\overline{e}})_\text{IO}&\simeq&x^2f^2\pm2xfg(y\sin\Delta-2\hat A |\epsilon_{e\mu}| \sin\phi_{e\mu}\cos\Delta)\,.
\label{eq:PIO}
\eea
From Eqs.~\eqref{eq:PNO} and~\eqref{eq:PIO}, we see that if $\epsilon_{ee}=0$ and $\phi_{e\mu}=0$ or $180^\circ$,
the differences \tc{between with and without NSIs in appearance probabilities} disappear in both the neutrino and antineutrino
modes, which are consistent with the degeneracy between $|\epsilon_{e\mu}|$ and $|\epsilon_{e \tau}|$ found in \cite{Liao:2016orc}. Note that for large $|\epsilon_{e\mu}|$,  higher order terms cannot be neglected,  and the degeneracy is broken. 

From Eq.~(\ref{eq:app-prob}), we see that $\epsilon_{\mu\tau}$ does not appear in the
appearance probability up to second order in $\epsilon$. Hence, they
mainly affect the disappearance channel. Taking $\epsilon_{ee}$,
$\epsilon_{e\mu}$ and $\epsilon_{e\tau}$ to zero, the disappearance
probability can be written as
\bea
P_{\mu \mu} &=& 1- \sin^2\Delta+ r c_{12}^2 \Delta\sin2\Delta-\frac{s_{13}^2\sin^2(1-\hat{A})\Delta}{(1-\hat{A})^2}
\nonumber\\
&-&\frac{s_{13}^2}{(1-\hat{A})^2}\left[\hat{A}(1-\hat{A})\Delta\sin2\Delta+\sin(1-\hat{A})\Delta\sin(1+\hat{A})\Delta\right]
\nonumber\\
&-& 2 \hat{A}|\epsilon_{\mu\tau}|\cos\phi_{\mu\tau}(\Delta\sin2\Delta)
-2\hat{A}^2 |\epsilon_{\mu\tau}|^2 \left(2\cos^2\phi_{\mu\tau}\Delta^2\cos2\Delta
+\sin^2\phi_{\mu\tau}\Delta  \sin2\Delta\right)
\nonumber\\
&+& {\cal O}(s_{13}^2\epsilon, r\epsilon, s_{13}\epsilon^2,  \epsilon^3)\,.
\label{eq:dis-prob}
\eea
The above equation shows that if $\phi_{\mu\tau}=90$ or $270^\circ$, the disappearance probability only depends on the second order in $\epsilon$. 

Finally, it is important to emphasize that $\tilde{\epsilon}_{ee}$ and $\epsilon_{e\mu}$ are the leading order for NSIs in the $\nu_\mu\rightarrow\nu_e$ appearance [c.f. Eq.~\eqref{eq:app-prob}] channel, and $\epsilon_{\mu\tau}$ also appears as the leading order in the  $\nu_\mu\rightarrow\nu_\mu$ disappearance [c.f. Eq.~\eqref{eq:dis-prob}] channel. Therefore, combining data of the appearance and disappearance channels can be used to measure all three NSI parameters in the NSI matrix under the $\mu-\tau$ reflection symmetry in Eq.~\eqref{eq:V}. As a result, we consider the planned LBL neutrino oscillation experiments that measure both $P_{\mu e}$ and $P_{\mu\mu}$ to test the $\mu-\tau$ reflection symmetry in the presence of NSIs.

\subsection{Experimental and simulation details}\label{sec:expt}
In this study, we consider two proposed next-generation superbeam experiments DUNE~\cite{Acciarri:2015uup,Alion:2016uaj}  and T2HK~\cite{Abe:2018uyc}. 
We propose combining these two experiments to test the $\mu-\tau$ reflection symmetry in NSIs.
The advantage of this configuration is that combining T2HK with DUNE can be used to reduce the impact of standard oscillation parameter uncertainties on the measurement of NSI parameters.
To perform the numerical simulation of both the experiments, we use the \texttt{GLoBES} packages \cite{Huber:2004ka,Huber:2007ji} along with the required auxiliary files presented in Ref.~\cite{Alion:2016uaj} for DUNE and the detector simulation files in Ref. \cite{Liao:2016orc} for T2HK. DUNE will utilize existing Neutrinos at the Main Injector beamline design at Fermilab as a neutrino source. The far detector of DUNE will be placed at Sanford Underground Research Facility in Lead, South Dakota, at a distance of 1300 km (800 mile) and about 1.5 km under the surface from the neutrino source. DUNE Collaboration has planned to use four 10 kton Liquid Argon Time-Projection Chamber detectors. According to their report, the expected design flux corresponds to 1.07 MW beam power which gives $1.47\times 10^{21} $ protons on target (POT) per year for an 80 GeV proton beam energy. In our simulation, we simply consider 40 kton of fiducial mass for the far detector. In addition, the signal and background normalization uncertainties for appearance and disappearance channels have been adopted from DUNE CDR~\cite{Alion:2016uaj}.  We perform our numerical study considering 3.5 years running time in both neutrino and antineutrino modes.  

The T2HK experiment~\cite{Abe:2018uyc} will utilize an upgraded J-PARC beam with a power of 1.3 MW, which gives $2.7\times 10^{21} $ POT per year for a 30 GeV proton beam energy. They are planning to build a Water Cherenkov detector, which is located 295 km away from the source. Making the detector $2.5^\circ$ off axis allows it to measure a narrow band beam with a peak energy around 0.6 GeV. Here we consider the 2TankHK-staged design proposed in the HK design report~\cite{Abe:2018uyc}. We assume one tank will take data alone for the first 6 years, whereas a second tank will be added for another 4 years. Each tank will have 0.19 Mton fiducial mass with 40\% photo coverage. While performing the numerical analysis, we take 2.5 years running time in neutrino modes and 7.5 years running time in antineutrino modes, i.e., 1 : 3 ratio out of the total runtime.

\begin{table}[]
	\begin{tabular}{|c|c|c|c|c|c|c|}
		\hline
		Parameter & $\theta_{12}/^\circ$ & $\theta_{13}/^\circ$ & $\theta_{23}/^\circ$ & $\delta/^\circ$  & $\Delta m_{21}^2/10^{-5}\text{eV}^2$ & $\Delta m_{31}^2/10^{-3}\text{eV}^2$\\ \hline
		Best fit & $33.82^{+0.78}_{-0.76}$ & $8.61^{+0.13}_{-0.13}$ & $49.6^{+1.0}_{-1.2}$ & $215^{+40}_{-29}$ & $7.39^{+0.21}_{-0.2}$ & $2.525^{+0.033}_{-0.032}$ \\\hline
		$3\sigma$ range & $31.61-36.27$ & $8.22-8.99$ & $40.3-52.4$ & $125-392$ & $6.79-8.01$ & $2.47-2.625$ \\\hline
	\end{tabular}
	\caption{The best fit and $1\sigma$ uncertainty, in the results of NuFit4.0 \cite{Esteban:2018azc} for normal mass ordering.}\label{tab:nufit4.0}
\end{table}

To incorporate NSIs, we use the GLoBES extension file snu.c as has been presented in Refs.~\cite{Kopp:2006wp,Kopp:2007ne}. To implement the $\mu-\tau$ reflection symmetry, we directly impose the form Eq.~\eqref{eq:V} on the matter potential in snu.c. We fix the true parameter values of NSIs at zero, whereas  the test parameter ranges of NSIs have been considered in~\cite{Ohlsson:2012kf}. 
Except for the value of $\delta$ and $\theta_{23}$, we use NuFit4.0 best-fit results for the true values~\cite{Esteban:2018azc}, as shown in Table~\ref{tab:nufit4.0}. We fix the value $\theta_{13}$ at the true value, as we have checked it does not make a great impact on the measurement on NSI parameters. We also fix $\theta_{12}$ and $\Delta m_{21}^2$ as the true values as these two parameters do not enter to the neutrino oscillations of DUNE and T2HK. 
We vary $\Delta m_{31}^2$ within a Gaussian prior with the true value and the width which are the best fit and the $1\sigma$ uncertainty of NuFit4.0 results. 
In addition, we consider two orderings: for normal (inverted) mass ordering, the central value and the $1\sigma$ relative width of the $\Delta m_{31}^2$ prior are $2.525$~eV$^2$ ($-2.586$~eV$^2$) and $1.3\%$ ($1.3\%$), respectively.  
We do not include any other priors. Furthermore, we vary the phases $\phi_{e\mu}$ and $\phi_{\mu\tau}$ in the range $[0,2\pi)$. Note that the $\mu-\tau$ reflection symmetry supports the phases $\phi_{e\mu}, \phi_{\mu\tau} \neq 0, \pi$.
Finally, we do not consider the LMA dark solution that is allowed by the generalized mass ordering degeneracy~\cite{Bakhti:2014pva,*Coloma:2016gei,*Deepthi:2016erc}, since it is difficult to achieve $\epsilon_{\alpha\beta} \sim \mathcal{O}(1)$ in the flavor symmetrical models.

\subsection{Constraints on NSI parameters}\label{sec:NSI_results}
\begin{figure}[h!]
\centering
   \includegraphics[width=8cm]{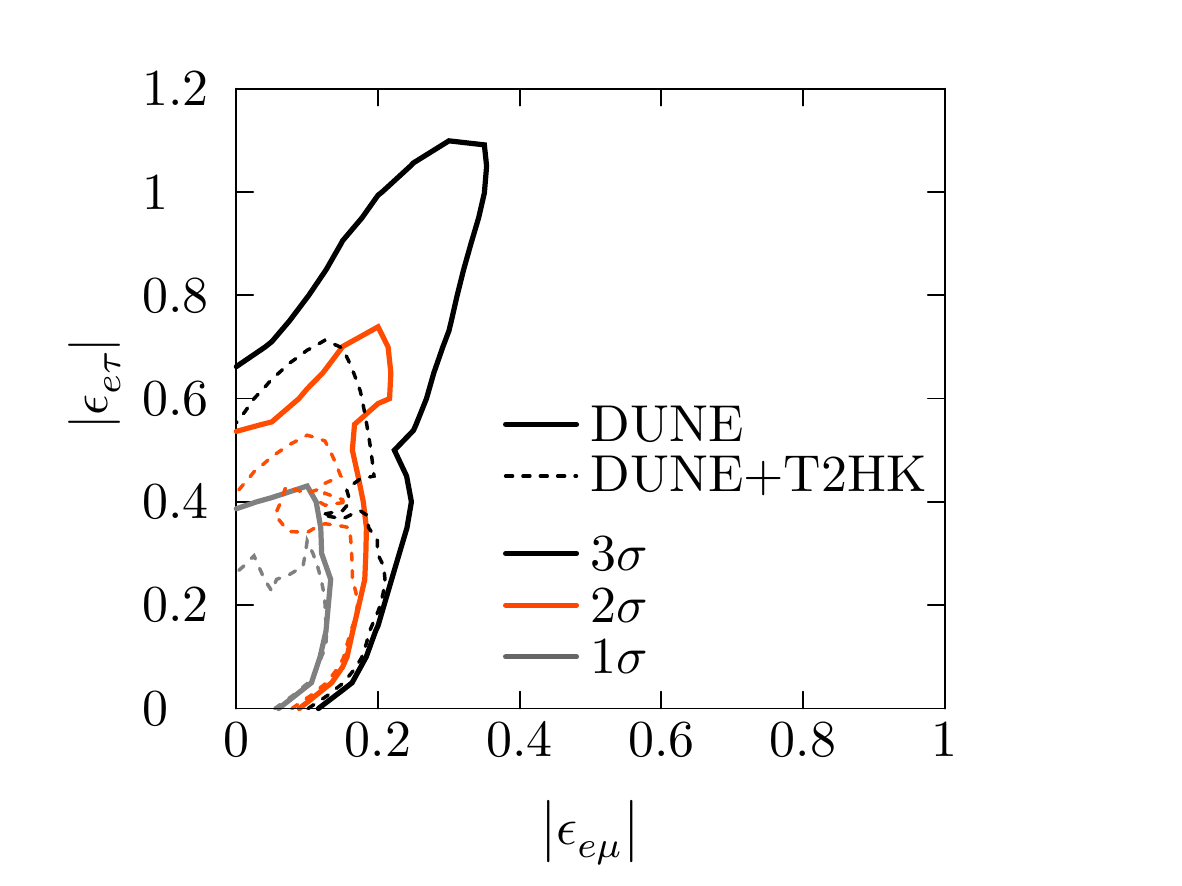}
\caption{The $1\sigma$ (gray), $2\sigma$ (red), and $3\sigma$ (black) contours on the $|\epsilon_{e\mu}| - |\epsilon_{e \tau}|$ plane for DUNE (solid) and the combination of DUNE and T2HK (dashed) without restricting the $\mu-\tau$ reflection symmetry on $\epsilon_{\alpha\beta}$. }
\label{fig:ContourEmuEtau}
\end{figure}

We first study the impact of imposing the $\mu-\tau$ reflection symmetry on the constraints on NSI parameters at next-generation LBL neutrino oscillation experiments.
In Fig.~\ref{fig:ContourEmuEtau}, we present the allowed parameter region on the $|\epsilon_{e\mu}|-|\epsilon_{e \tau}|$ plane without restricting the $\mu-\tau$ reflection symmetry on $\epsilon_{\alpha\beta}$ for DUNE (solid) and the synergy of DUNE and T2HK (dashed). Gray, orange, and black curves represent $1\sigma$, $2\sigma$, and $3\sigma $ contours, respectively.  The tendency of contours leaves away from $|\epsilon_{e\mu}|=|\epsilon_{e\tau}|$, because of the contribution from the higher-order term in $P(\nu_\mu\rightarrow\nu_e)$ and $P(\bar{\nu}_\mu\rightarrow\bar{\nu}_e)$. Along with $|\epsilon_{e\mu}|=|\epsilon_{e\tau}|$, the bounds at $1\sigma$, $2\sigma$, and $3\sigma$ are $\sim0.125$, $\sim0.16$, and $\sim0.19$, respectively. Obviously, this shows that when we impose the restriction of the $\mu-\tau$ reflection symmetry on NSI, the uncertainty of $|\epsilon_{e\mu}|$ and $|\epsilon_{e\tau}|$ is much reduced. We see a deficit for all contours. For example, for the $3\sigma$ contour for DUNE, the discontinuity appears around $|\epsilon_{e\mu}|\sim 0.25$ and $|\epsilon_{e\tau}|\sim 0.5$. This is due to the flipping of the sign of $\Delta m_{31}^2$.

\begin{figure}[!t]%
 \includegraphics[width=3in]{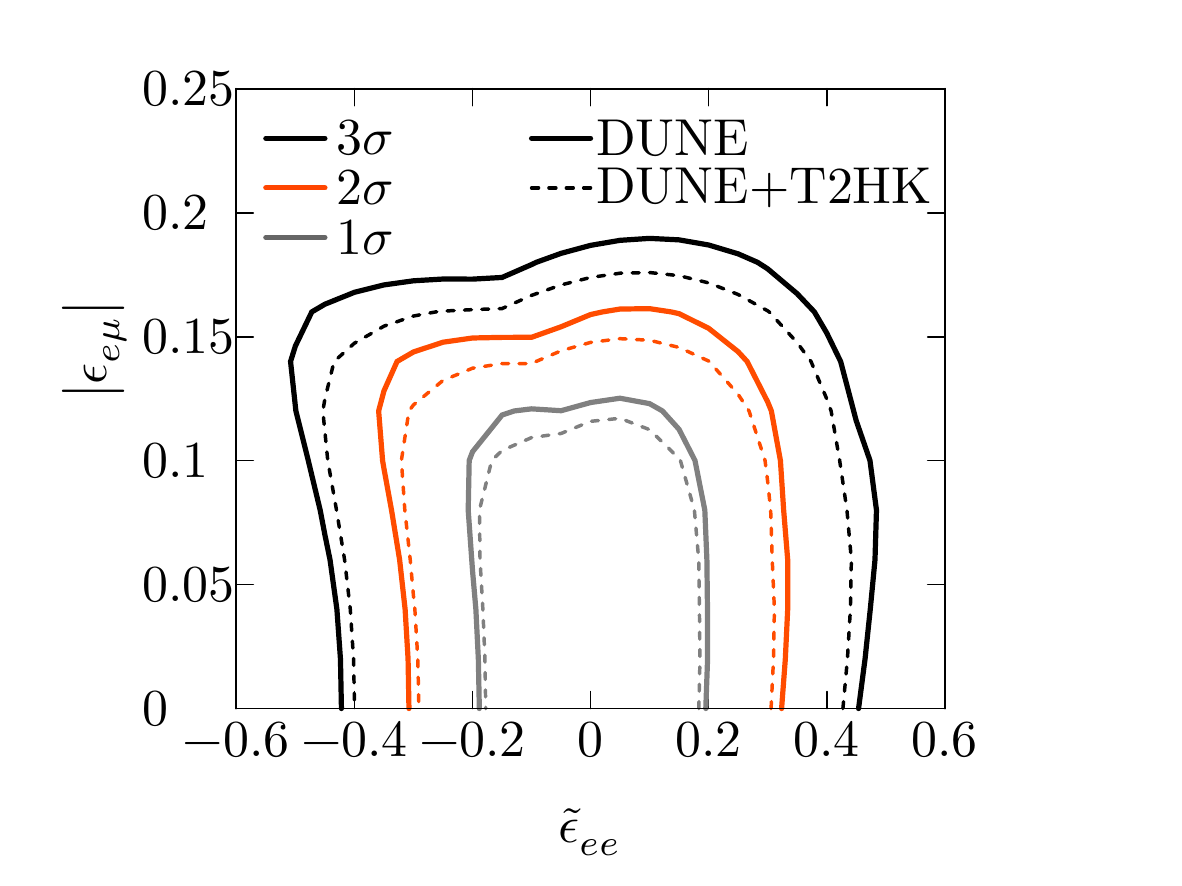}\hspace{-12mm}
  \includegraphics[width=3in]{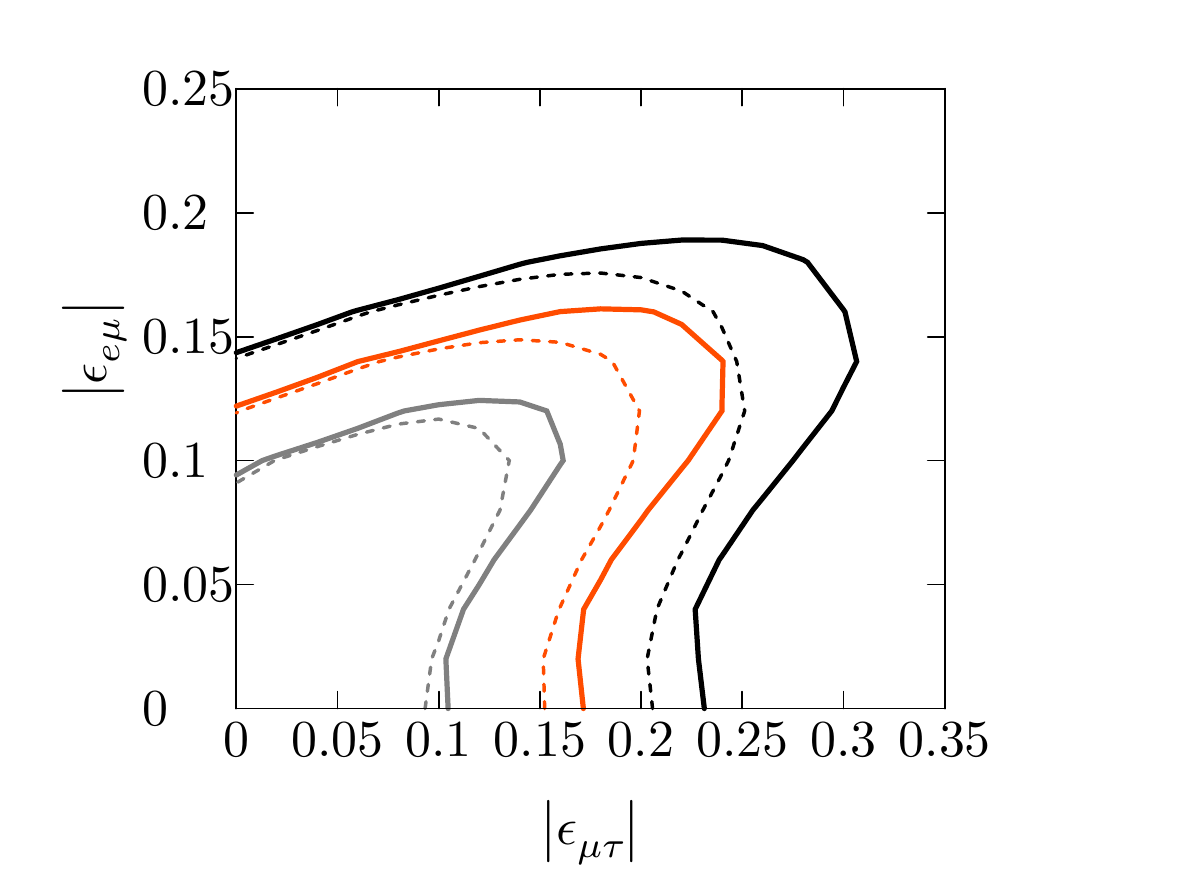}
\caption{The $1\sigma$ (gray), $2\sigma$ (red), and $3\sigma$ (black) contours on $\tilde{\epsilon}_{ee}-|\epsilon_{e\mu}|$ (upper) and $|\epsilon_{\mu\tau}|-|\epsilon_{e\mu}|$ (upper) plane for DUNE (solid) and the combination of DUNE and T2HK (dashed) with the $\mu-\tau$ reflection symmetry. }%
 \label{fig:epsilon}
\end{figure}

\begin{table}[]
\begin{tabular}{|c|c|c|}
\hline
Parameters$~~~~~~~~~$ & $1\sigma$ w/o $\mu-\tau$~~~~~ & $1\sigma$ w/ $\mu-\tau$\\\hline
$\tilde{\epsilon}_{ee}$$~~~~~~~~~$ & $[-2.5,~1.2]$$~~~~~$ & $[-0.13,0.13]$\\\hline
$|\epsilon_{e\mu}|$$~~~~~~~~~$ & $[0,~0.12]$$~~~~~$ & $[0,~0.1]$\\\hline
$|\epsilon_{e\tau}|$$~~~~~~~~~$ & $[0,~0.3]$$~~~~~$ & $[0,~0.1]$\\\hline
$|\epsilon_{\mu\tau}|$$~~~~~~~~~$ & $[0,~0.2]$$~~~~~$ & $[0,~0.12]$\\\hline
\end{tabular}
\caption{The comparison of the constraints on NSI parameters between cases with (w/) and without (w/o) restricting the $\mu-\tau$ reflection symmetry on $\epsilon_{\alpha\beta}$ for DUNE.}\label{tab:compare_DUNE}
\end{table}

In Fig.~\ref{fig:epsilon}, we discuss on the scenario in the presence of the $\mu-\tau$ reflection symmetry on NSI. We show $1\sigma$ (gray), $2\sigma$ (red), and $3\sigma$ (black) contours on $\tilde{\epsilon}_{ee}-|\epsilon_{e\mu}|$ (left), $|\epsilon_{\mu\tau}|-|\epsilon_{e\mu}|$ (right) planes for DUNE (solid) and the combination of DUNE and T2HK (dashed). We see a correlation between $|\epsilon_{e\mu}|$ and $|\epsilon_{\mu\tau}|$. This is because of the flipping of $\mathrm{sign}(\Delta m_{31}^2)$. In the other panel, we see that there is a dip on top of solid curves. This feature can be removed by including T2HK data. Talking about the size of constraints, though T2HK is not sensitive for measuring NSI parameters, including its data significantly improves the $|\epsilon_{\mu\tau}|$ measurement, for which T2HK data can reduce the size of the uncertainty by $\sim 20\%$. The improvement on the other-parameter measurements is weaker. We note that the size of uncertainties is reduced once the $\mu-\tau$ reflection symmetry is preserved in NSI. We compare the $1\sigma$ constraints between scenarios with and without the $\mu-\tau$ reflection symmetry for DUNE in Table~\ref{tab:compare_DUNE}. We see the improvement of the precision of $\tilde{\epsilon}_{ee}$ and $|\epsilon_{\mu\tau}|$ measurements, which is not only because of the fact that the number of degrees of freedom is reduced by imposing the $\mu-\tau$ reflection symmetry in the neutrino-mass matrix and NSI, but also the size of the allowed range for $|\epsilon_{e\tau}|$ is largely shrunken down.
As we see from Table~\ref{tab:compare_DUNE}, the allowed ranges for the NSI parameters with the $\mu-\tau$ reflection symmetry are in general smaller than those without the $\mu-\tau$ reflection symmetry. Therefore, it is more difficult to rule out the presence of NSIs with the $\mu-\tau$ reflection symmetry than that without the $\mu-\tau$ reflection symmetry. If DUNE measured NSI parameters that were larger than the values in the third column in Table~\ref{tab:compare_DUNE} but smaller than those in the second column in Table~\ref{tab:compare_DUNE}, it could be an indication of the presence of the $\mu-\tau$ reflection symmetry in the NSI matrix.

\subsection{Impact on neutrino mass ordering sensitivity}\label{sec:results}

The main goals of next-generation neutrino oscillation experiments are to measure the Dirac CP phase $\delta$, to determine the neutrino mass ordering, and the octant of $\theta_{23}$~\cite{Acciarri:2015uup}. Since the $\mu-\tau$ reflection symmetry is preserved in both the neutrino mass matrix and the NSI matrix in our formalism, $\theta_{23}=45^\circ$ and $\delta=\pm 90^\circ$ will be measured by future neutrino oscillation experiments even in the presence of NSI. We find that the wrong sign of $\delta$ can be excluded at more than $5\sigma$ C.L.~at DUNE and T2HK. Hence, here we fix  $\delta=- 90^\circ$ which is favored by the latest global analysis of neutrino oscillation data~\cite{Capozzi:2016rtj,*Esteban:2016qun,*deSalas:2017kay,*Esteban:2018azc}, and only focus on the study of the impact of NSI on the determination of mass ordering sensitivity at DUNE and T2HK within this formalism.

\begin{figure}[tb]
\centering
\includegraphics[width=12cm, height=8cm]{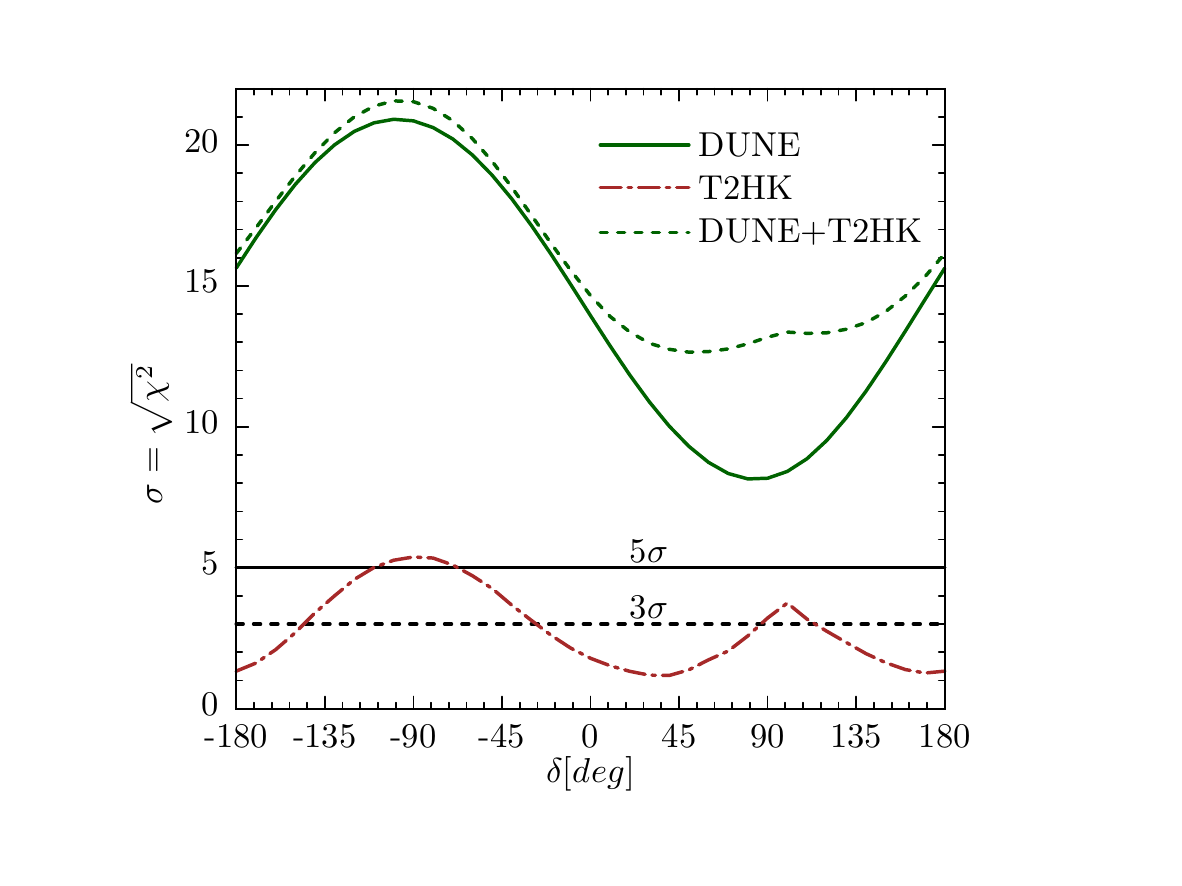}
   \vspace{-8mm}
\caption{Mass ordering sensitivity in the presence of standard matter interactions. Here, green solid (brown dashed-dotted) curve represents the $ \chi^{2} $ sensitivity for DUNE (T2HK), whereas their combined analysis are shown by the green dotted curve. }
\label{fig:MH-SI}
\end{figure}

We first examine the potential of both experiments to measure the mass ordering in the presence of standard matter interactions as shown in Fig.~\ref{fig:MH-SI}. The combined potential of both experiments is also investigated. The solid green curve shows the ordering $ \chi^{2} $ for DUNE, and the dashed-dotted brown curve represents the same for T2HK. The combined potential of DUNE and T2HK are shown by the dotted green curve.
We also mark the $3\sigma$, $5\sigma $ significance levels by the solid and dotted black horizontal lines, respectively. Our result is consistent with that in Ref.~\cite{Ballett:2016daj}. We notice that DUNE and T2HK can achieve a maximum mass ordering sensitivity of around $ \sqrt{\chi^{2}} \sim 20$ and $5$ significance levels, around the true value of $ \delta = - 90^\circ $, respectively. Though the mass ordering sensitivity for T2HK is not good as DUNE, we see an improvement by combining two sets of data. Thus, in what follows we concentrate  about the sensitivities of DUNE and the combination of DUNE+T2HK.

\begin{figure}[tb]
\centering
   \includegraphics[width=8cm, height=6cm]{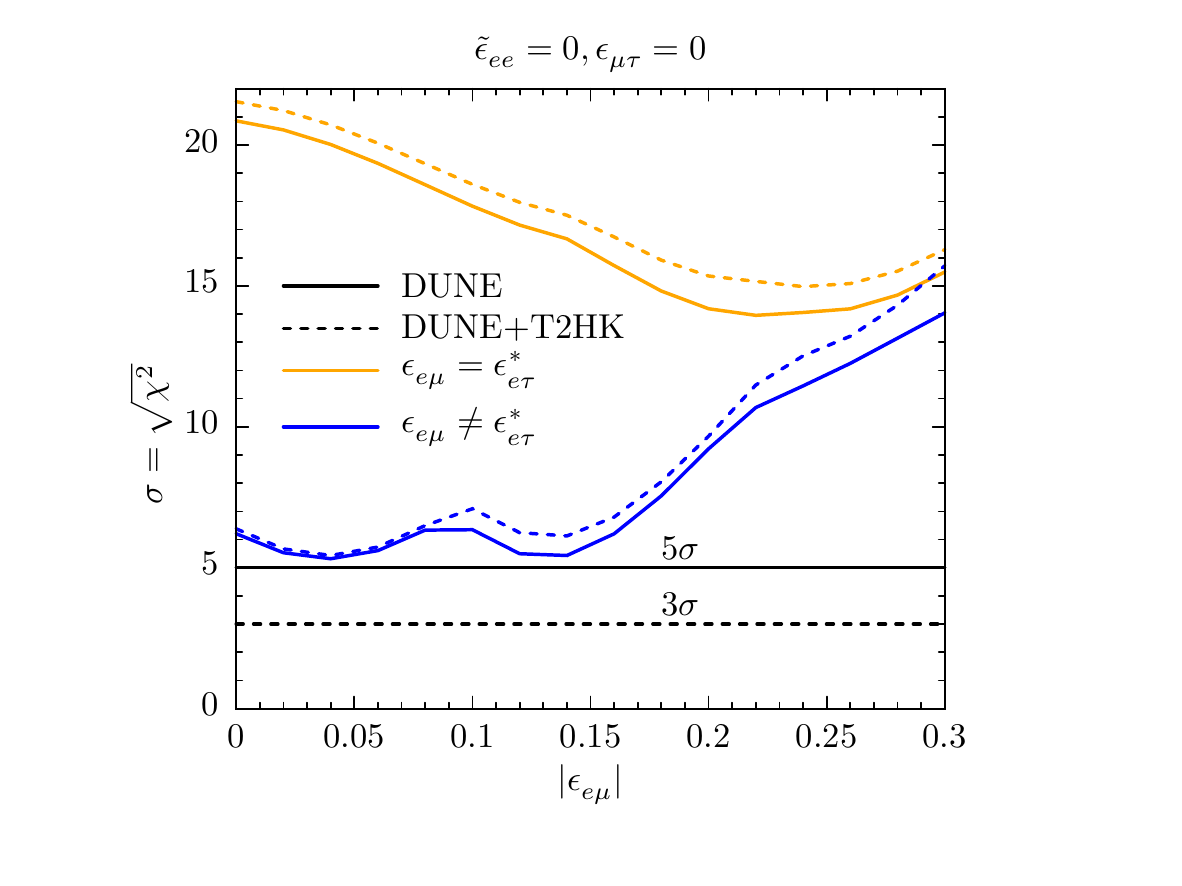}\hspace{-12mm}
\includegraphics[width=8cm, height=6cm]{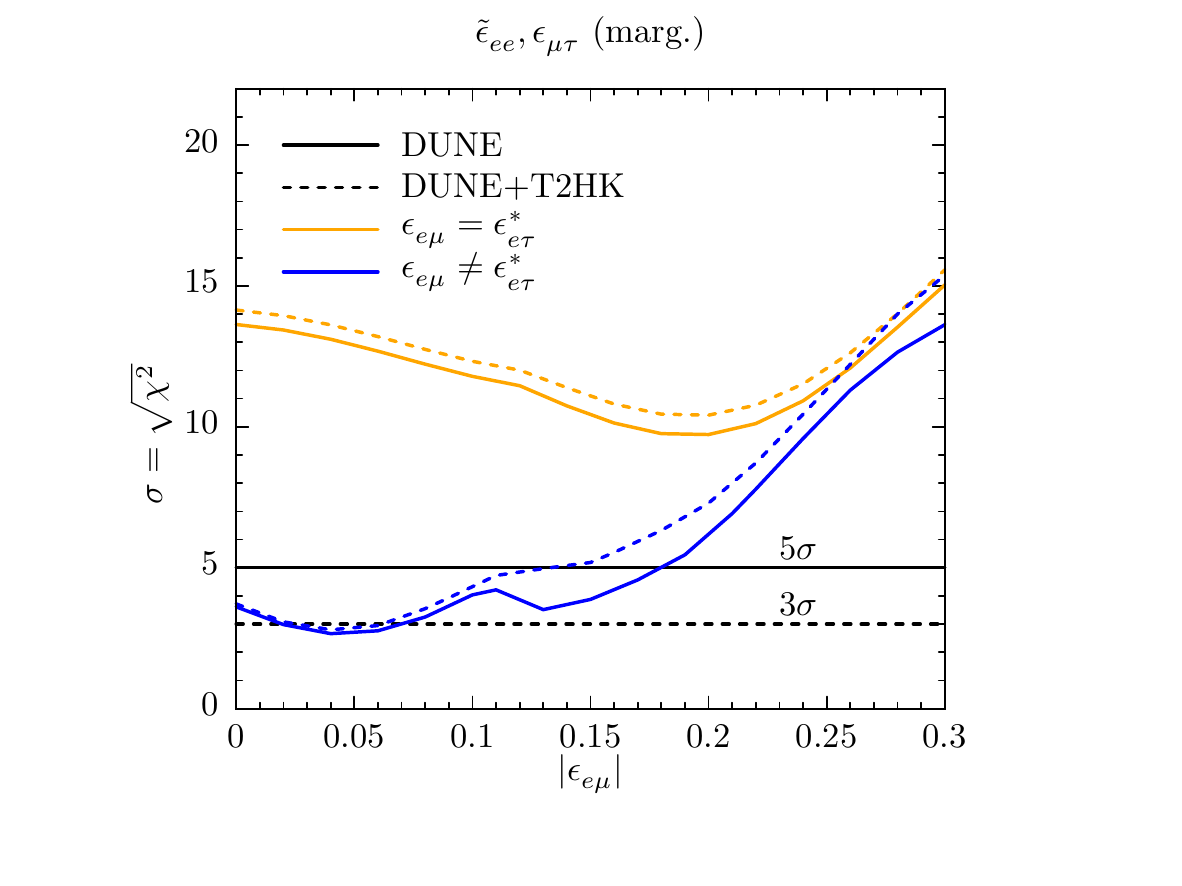}
\vspace{-8mm}
\caption{ Left: mass ordering sensitivity  vs  $\epsilon_{e\mu}$ for  $\tilde{\epsilon}_{ee} = 0, \epsilon_{\mu \tau} = 0$ and $ \delta = - 90^\circ $, where solid (dashed) curves show the sensitivity level for DUNE (the combination of DUNE and T2HK). Also, the orange (blue) curves depict our results with (without) the $\mu-\tau$ reflection symmetry on $\epsilon_{\alpha\beta}$. Right: same as the left, except for values of $\tilde{\epsilon}_{ee}$ and $\epsilon_{\mu \tau}$ marginalized over (marg.).}
\label{fig:MassHierEE0}
\end{figure}

To investigate the impact of  NSI on the determination of neutrino mass ordering under the adopted formalism, we present our results in Fig.~\ref{fig:MassHierEE0}. For simplicity, we first discuss our results for nonzero $\epsilon_{e\mu}, \epsilon_{e\tau}$, where the remaining NSIs (i.e.,  $\tilde{\epsilon}_{ee}, \epsilon_{\mu \tau}$) have been fixed at zero. It has been presented in the left panel of Fig.~\ref{fig:MassHierEE0}.  We show our results for the DUNE (DUNE+T2HK) using the solid curves (the dotted curves). The orange curves show our results in presence of   the symmetry, whereas the blue curves represent the same, but in the absence of the symmetry. 
We emphasize here that while performing our numerical simulation, NSI parameters $ \epsilon_{e\mu},  \epsilon_{e\tau}$ along with their respective phases $ \phi_{e\mu}, \phi_{e\tau} $ have been marginalized for the left panel. In addition, we marginalize $\tilde{\epsilon}_{ee}, \epsilon_{\mu \tau}$,  and $   \phi_{\mu\tau}$ for the right panel. 

%
In case of DUNE, we notice that the mass ordering sensitivity does not get much affected,
when the NSI parameters respect the symmetry over the whole parameter space (see solid orange curve of the left panel of Fig.~\ref{fig:MassHierEE0}). 
However, the sizable impact has been observed for the scenario where NSI parameters do not respect the symmetry (i.e. for $\epsilon_{e\mu} \neq \epsilon^*_{e\tau}$), which we show by the solid blue curve for the DUNE (see figure levels for details). The sensitivity curve gets severely  affected for $ \epsilon_{e\mu} \lesssim 0.15 $ as can be seen from the solid blue curve.
The combined analysis of DUNE+T2HK does not show much improvement in the sensitivity, as seen from the dotted blue curve. In the right panel of Fig.~\ref{fig:MassHierEE0}, we also demonstrate the mass ordering $ \chi^{2} $ similar to the left panel, but for nonzero $\tilde{\epsilon}_{ee}$, $\epsilon_{e\mu}, $ and $ \epsilon_{\mu\tau}$.
A similar pattern on the mass ordering sensitivity has been identified. 
We notice from the blue curves that the sensitivity can come down below $ 5\sigma $ for $\epsilon_{e\mu} \neq \epsilon^*_{e\tau}$.
However, more than $ 10\sigma $ sensitivity has been observed for $\epsilon_{e\mu} = \epsilon^*_{e\tau}$, as shown by the orange curves.

\section{Summary\label{sec:summary}}

The $\mu-\tau$ reflection symmetry, which predicts the maximal value of the atmospheric mixing angle, and the Dirac CP-violating phase are in well agreement with the latest neutrino oscillation data. In this work, we explore the possibility of embedding the $\mu-\tau$ reflection symmetry in the NSI matrix and study its importance for the determination of neutrino mass ordering in the case of  DUNE and T2HK. 

We present an $ S_4 $ flavor model, with the $\mu-\tau$ reflection symmetry realized in both neutrino mass and NSI matrixes. In this model, the sizable NSI effects originate from the mediator of scalar-and-doublet-singlet mixing. 
The essential to achieve the $\mu-\tau$ reflection symmetry is the introduction of flavon fields with vacuum satisfying the symmetry. By coupling the flavons to neutrinos,  the $\mu-\tau$ reflection symmetry in the neutrino mass matrix is realized. By coupling the flavons to the scalar mediators, \tc{this symmetry} in the NSI matrix is achieved. The model provides  a very generic way to construct NSIs with the $\mu-\tau$ reflection symmetry. 

We find that imposing the $\mu-\tau$ reflection symmetry improves the constraints on NSI parameters. In particular, the constraint on $\tilde{\epsilon}_{ee}$ is improved most; its allowed regime is improved from $[-2.5,~1.2]$ (without the $\mu-\tau$ reflection symmetry) to $[-0.13,~0.13]$ (within the $\mu-\tau$ reflection symmetry). It is not only because of the reduced numbers of NSI parameters, but also due to the fact that the allowed region for $\epsilon_{e\tau}$ is reduced. This exactly points out that NSIs under this scenario is much more testable than the generalized case, with the upcoming neutrino oscillation experiments.

We also study the impact of NSIs on the mass ordering sensitivity. Assuming $ \tilde{\epsilon}_{ee} = \epsilon_{\mu \tau}=0$, we find that in the presence of the $\mu-\tau$ reflection symmetry, DUNE alone can reach almost 14$ \sigma $ mass ordering sensitivity. However,  the mass ordering sensitivity gets compromised in the absence of the concerned symmetry as presented in  Fig.~\ref{fig:MassHierEE0}. We find slightly better sensitivity when combining DUNE and T2HK. Considering the most general scenario under the $\mu-\tau$ reflection symmetry, we find that DUNE  can achieve more than 10$ \sigma$ sensitivity, whereas the combined analysis of DUNE and T2HK data can give relatively better sensitivity than DUNE.

In summary, the existence of the $\mu-\tau$ reflection symmetry in the neutrino mass matrix and matter effect NSIs is attractive in both theoretical and phenomenological points of view. It is naturally realized in the framework of flavor symmetries and is consistent with the current data. Furthermore, once the $\mu-\tau$ reflection symmetry is the true symmetry behind, the testability of NSIs is enhanced, and the impact of NSIs on the mass ordering sensitivity is reduced.

\acknowledgments
NN thanks Prof. Zhi-zhong Xing for an opportunity to visit Sun Yat-sen University, Guangzhou and also acknowledges their warm hospitality, where the work has been initiated.  
JL is supported in part by National Natural Science Foundation of China under Grant No.~11905299.
NN is supported in part by the National Natural Science Foundation of China under Grant No.~11775231.
TCW is supported in part by the National Natural Science Foundation of China under Grant No.~11505301 and No.~11881240247. 
YLZ acknowledges the STFC Consolidated Grant ST/L000296/1 and the European Union's Horizon 2020 Research and Innovation programme under Marie Sk\l{}odowska-Curie grant agreements Elusives ITN No.\ 674896 and InvisiblesPlus RISE No.\ 690575.

\appendix

\section{Group theory of $S_4$ \label{app:S4}} 

$S_4$ is the permutation group of four objects. Its three generators $S$, $T$, and $U$ satisfying the equalities $T^3=S^2=U^2=(ST)^3=(SU)^2=(TU)^2=1$, from which $(STU)^4=1$ is automatically obtained. The minimal number of generators of $S_4$ is actually two \cite{Bazzocchi:2009pv, Ding:2013eca}. However, we follow the presentation in \cite{deMedeirosVarzielas:2017hen} to emphasize the $Z_2$ residual symmetries generated by $S$ and $U$, respectively. $S_4$ contains five irreducible representations (irreps), $\mathbf{1}$, $\mathbf{1}'$, $\mathbf{2}$, $\mathbf{3}$, and $\mathbf{3}'$. The Kronecker products between different irreps can be easily obtained,
\begin{eqnarray}
&&\hspace{-1cm}
\mathbf{1}\otimes\mathbf{r}=\mathbf{r},~
\mathbf{1^{\prime}}\otimes\mathbf{1^{\prime}}=\mathbf{1}, ~
\mathbf{1^{\prime}}\otimes\mathbf{2}=\mathbf{2}, ~
\mathbf{1^{\prime}}\otimes\mathbf{3}=\mathbf{3^{\prime}}, ~
\mathbf{1^{\prime}}\otimes\mathbf{3^{\prime}}=\mathbf{3},~
\mathbf{2}\otimes\mathbf{2}=\mathbf{1}\oplus\mathbf{1}^{\prime}\oplus\mathbf{2},\nonumber\\
&&\hspace{-1cm}
\mathbf{2}\otimes\mathbf{3}=\mathbf{2}\otimes\mathbf{3^{\prime}}=\mathbf{3}\oplus\mathbf{3}^{\prime},\,
\mathbf{3}\otimes\mathbf{3}=\mathbf{3^{\prime}}\otimes\mathbf{3^{\prime}}=\mathbf{1}\oplus \mathbf{2}\oplus\mathbf{3}\oplus\mathbf{3^{\prime}},\,
\mathbf{3}\otimes\mathbf{3^{\prime}}=\mathbf{1^{\prime}}\oplus \mathbf{2}\oplus\mathbf{3}\oplus\mathbf{3^{\prime}},
\end{eqnarray}
where $\mathbf{r}$ represents any irrep of $S_4$. 
Throughout this paper, we work in the basis where the generator $T$ is diagonal. Generators of $S_4$ in different irreps are listed in Table \ref{tab:rep_matrix2}. In this basis, once we arrange $a= (a_1, a_2, a_3)^T$ as a triplet $\mathbf{3}$ (or $\mathbf{3}'$) of $S_4$, $\tilde{a} = (a_1^*, a_3^*, a_2^*)^T$, instead of $(a_1^*, a_2^*, a_3^*)^T$, transforms as a triplet of $S_4$. 

\begin{table}[h!]
\begin{center}
\begin{tabular}{cccc}
\hline\hline
   & $T$ & $S$ & $U$  \\\hline
$\mathbf{1}$ & 1 & 1 & 1 \\
$\mathbf{1^{\prime}}$ & 1 & 1 & $-1$ \\
$\mathbf{2}$ & 
$\left(
\begin{array}{cc}
 \omega  & 0 \\
 0 & \omega ^2 \\
\end{array}
\right)$ & 
$\left(
\begin{array}{cc}
 1 & 0 \\
 0 & 1 \\
\end{array}
\right)$ & 
$\left(
\begin{array}{cc}
 0 & 1 \\
 1 & 0 \\
\end{array}
\right)$ \\

$\mathbf{3}$ &  $\left(
\begin{array}{ccc}
 1 & 0 & 0 \\
 0 & \omega ^2 & 0 \\
 0 & 0 & \omega  \\
\end{array}
\right)$ &
$\frac{1}{3} \left(
\begin{array}{ccc}
 -1 & 2 & 2 \\
 2 & -1 & 2 \\
 2 & 2 & -1 \\
\end{array}
\right)$ &
$\left(
\begin{array}{ccc}
 1 & 0 & 0 \\
 0 & 0 & 1 \\
 0 & 1 & 0 \\
\end{array}
\right)$ \\

$\mathbf{3^{\prime}}$ &  $\left(
\begin{array}{ccc}
 1 & 0 & 0 \\
 0 & \omega ^2 & 0 \\
 0 & 0 & \omega  \\
\end{array}
\right)$ &
$\frac{1}{3} \left(
\begin{array}{ccc}
 -1 & 2 & 2 \\
 2 & -1 & 2 \\
 2 & 2 & -1 \\
\end{array}
\right)$ &
$-\left(
\begin{array}{ccc}
 1 & 0 & 0 \\
 0 & 0 & 1 \\
 0 & 1 & 0 \\
\end{array}
\right)$ \\ \hline\hline

\end{tabular}
\caption{\label{tab:rep_matrix2} The representation matrices for the $S_4$ generators $T$, $S$ and $U$, where $\omega=e^{2\pi i/3}$.}
\end{center}
\end{table}

Given two triplets $a= (a_1, a_2, a_3)^T$ and $b= (b_1, b_2, b_3)^T$, irreps for products of $a$ and $b$ are given by
\begin{eqnarray}
(ab)_\mathbf{1} &=& a_1b_1 + a_2b_3 + a_3b_2 \,,\nonumber\\
(ab)_\mathbf{2} &=& (a_2b_2 + a_1b_3 + a_3b_1,~ a_3b_3 + a_1b_2 + a_2b_1)^T \,,\nonumber\\
(ab)_{\mathbf{3}} &=& (2a_1b_1-a_2b_3-a_3b_2, 2a_3b_3-a_1b_2-a_2b_1, 2a_2b_2-a_3b_1-a_1b_3)^T \,,\nonumber\\
(ab)_{\mathbf{3}'} &=& (a_2b_3-a_3b_2, a_1b_2-a_2b_1, a_3b_1-a_1b_3)^T \,,
\label{eq:CG1}
\end{eqnarray}
for $a, b\sim \mathbf{3}$ or  $a, b\sim \mathbf{3}'$, and
\begin{eqnarray}
(ab)_{\mathbf{1}'} &=& a_1b_1 + a_2b_3 + a_3b_2 \,,\nonumber\\
(ab)_\mathbf{2} &=& (a_2b_2 + a_1b_3 + a_3b_1,~ -(a_3b_3 + a_1b_2 + a_2b_1))^T \,,\nonumber\\
(ab)_{\mathbf{3}'} &=& (2a_1b_1-a_2b_3-a_3b_2, 2a_3b_3-a_1b_2-a_2b_1, 2a_2b_2-a_3b_1-a_1b_3)^T \,,\nonumber\\
(ab)_{\mathbf{3}} &=& (a_2b_3-a_3b_2, a_1b_2-a_2b_1, a_3b_1-a_1b_3)^T \,,
\label{eq:CG2}
\end{eqnarray}
for $a \sim \mathbf{3}$ or  $b\sim \mathbf{3}'$. Contractions of a triplet $a= (a_1, a_2, a_3)^T \sim \mathbf{3}'$ and doublet $b= (b_1, b_2)^T\sim \mathbf{2}$ are given by
\begin{eqnarray}
(ab)_{\mathbf{3}'} &=& (a_2b_1+a_3b_2, a_3b_1+a_1b_2, a_1b_1+a_2b_2)^T \,,\nonumber\\
(ab)_{\mathbf{3}} &=& (a_2b_1-a_3b_2, a_3b_1-a_1b_2, a_1b_1-a_2b_2)^T \,.
\label{eq:CG2}
\end{eqnarray}

Lagrangian terms contributing to lepton masses and invariant under $S_4 \times Z_4$ are given by
\begin{eqnarray} \label{eq:mass_term}
    -\mathcal{L} &\supset&
    \frac{y_{e1}}{\Lambda^3} 
    \left( \overline{L} \left( \varphi (\varphi\varphi)_{\mathbf{2}} \right)_{\mathbf{3}}\right)_{\mathbf{1}} He_R + \frac{y_{e2}}{\Lambda^3} 
    \left( \overline{L} \left( \varphi (\varphi\varphi)_{\mathbf{3}} \right)_{\mathbf{3}}\right)_{\mathbf{1}} He_R +
     \frac{y_\mu}{2\Lambda^2} 
     \left( \overline{L} (\varphi\varphi)_{\mathbf{3}} \right)_{\mathbf{1}} H\mu_R 
      \nonumber\\
     &+& 
     \frac{y_\tau}{\Lambda} 
     \left( \overline{L} \varphi \right)_{\mathbf{1}'} H \tau_R
    + \frac{\tilde{H}\tilde{H}}{2\Lambda_{\rm ss} \Lambda} 
    \left[ 
    y_1 \left( \overline{L} L^c \right)_{\mathbf{1}} \xi +
    y_2 \left( \left( \overline{L} L^c \right)_{\mathbf{2}} \zeta \right)_{\mathbf{2}} + 
    y_3 \left( \left( \overline{L} L^c \right)_{\mathbf{3}} \chi \right)_{\mathbf{3}}
    \right] + {\rm H.c.}.\,\,\,\,\,\,\,\,\,\,\,\,\,
\end{eqnarray}
Applying the Clebsch-Gordan (CG) coefficients to the above formula, we arrive at  Eq.~\eqref{eq:mass_term} with $y_e = -y_{e1}+2y_{e2}$. 
With the help of the CG coefficients, Lagrangian terms contributing to NSI in Eq.~\eqref{eq:charged_scalar_Lagrangian} can be written in a more compact form,
\begin{eqnarray} \label{eq:charged_scalar_Lagrangian_2}
    -\mathcal{L} &\supset& \mu_\eta^2 (\tilde{\eta} \eta)_{\mathbf{1}} + \mu_\phi^2 \phi^- \phi^+ + \left( f \phi^- \tilde{H}^\dag (\eta \chi)_{\mathbf{1}}  + {\rm H.c.}\right) +\lambda 
    (\overline{L} \eta)_{\mathbf{1}}
    e_R \,,
\end{eqnarray}
where $\tilde{\eta} = (\eta_1^\dag, \eta_3^\dag, \eta_2^\dag)$ is understood. They are invariant under the symmetry $S_4 \times Z_4 \times Z_2$.

\bibliographystyle{apsrev4-1}
\bibliography{mu-tau.bib}
\end{document}